\documentclass[prb,twocolumn,aps,amssymb,footinbib,showpacs,floatfix]{revtex4-1}
\usepackage{amssymb}
\usepackage{graphicx}
\usepackage{amsmath}
\usepackage{times}
\usepackage{color}
\usepackage{subfigure}
\usepackage{setspace}
\usepackage{bm}% bold math

\newcommand {\kk} {\ensuremath{{\bf k}}}
\newcommand {\upp} {\ensuremath{{|\!\!\uparrow\rangle}}}
\newcommand {\dnn} {\ensuremath{{|\!\!\downarrow\rangle}}}

\begin{document}
\begin{abstract}
We consider a localized impurity atom that interacts with a cloud of fermions in the paired state. We develop an effective scattering length description of the interaction between an impurity and a fermionic atom using their vacuum scattering length. Treating the pairing of fermions at the mean-field level, we show that the impurity atom acts like a magnetic impurity in the condensed matter context, and leads to the formation of a pair of Shiba bound states inside the superconducting gap. In addition, the impurity atom can lead to the formation of deeply bound states below the Fermi sea. 
\end{abstract}

\title{Bound states of a localized magnetic impurity in a superfluid of paired ultracold fermions}
\author{Eric Vernier$^{1,2}$, David Pekker$^{1}$, Martin W. Zwierlein$^{3}$, Eugene Demler$^{1}$}

\affiliation{
$^1$ Physics Department, Harvard University, Cambridge, Massachusetts 02138, USA\\
$^2$ D\'epartement de Physique, Ecole Normale Sup\`erieure, Paris, France \\
$^3$ MIT-Harvard Center for Ultracold Atoms, Research Laboratory of Electronics,
and Department of Physics, Cambridge, MA 02139, USA
}

\pacs{67.85.-d, 67.85.Lm, 37.10.Jk}

\maketitle

Magnetic impurities in superconductors are known not only to alter the BCS ground-state by introducing potential scattering, but also to be at the origin of the pair-breaking effect leading to elementary excitations fundamentally different than those present in pure superconductors. Their presence serves to attenuate superconductivity by formation of in-gap Shiba bound states~\cite{Shiba}. In fact, at high concentrations magnetic impurities induce gapless superconductivity~\cite{AGD}. While magnetic impurities in a superconductor have often been discussed as one of the simplest models that exhibit interplay between superconductivity and magnetism, the consequences of this interplay are still not fully understood.

The experimental realization of paired states in ultracold atom systems has shed new light on many problems in superconductivity. In particular it has been instrumental in prompting a better understanding of the BEC-BCS crossover~\cite{Popov, Keldysh, Eagles, Leggett, Jin, Bartenstein,Zwierlein2004, Kinast, Bourdel, Zwierlein2005, Burovski,  Stringari}, including the case of pairing in systems with large spin imbalance~\cite{Ketterle, Hulet}. It has also prompted a new generation of research on the subject of the dynamics of theses systems~\cite{Barankov, Kehrein}. 

Combine magnetic impurities and ultracold atom systems can have rich physical consequences, some of which we explore in the present paper. In particular, introducing magnetic impurities into fermionic superfluids would help in understanding the interactions between magnetism and superfluidity, and could help to resolve long standing problems such as how superconductivity becomes gapless. 

Alternatively, instead of studying how the system changes in response to magnetism, one can use localized magnetic impurities as a form of local probe. As an example, in the setting of high temperature superconductors, detection of the modulation of the local density of states by a magnetic impurity via a scanning tunneling microscopy (STM) was used to great advantage to probe the nature of quasi-particle states of these materials both in the superconducting and pseudo-gap phases~\cite{Davis, BalatskyViews, Hoffman}. 

Although STM spectroscopy is not currently possible in the ultracold atom setting Radio Frequency (RF) spectroscopy~\cite{Spectroscopy, FermiPolaron} could be used to probe the nature of the superconducting state in the vicinity of the magnetic impurity. Combining a magnetic impurity with RF spectroscopy can be used to directly probe the size of the superconducting gap eliminating the uncertainty due to effects like Hartree shifts~\cite{Baym, Schirotzek}. Further, we envision that additional information from momentum resolved RF spectroscopy~\cite{Stewart} in the vicinity of a magnetic impurity could provide data on the symmetry of the gap and its nodal structure. 

In this paper we propose and investigate theoretically a scheme for introducing localized magnetic impurities into the ultracold atom fermionic superfluid. The impurity is formed by an atom of a different  species (from the species making up the superfluid) that is localized by a deep optical lattice potential. The laser frequency is chosen such that the optical lattice interacts only weakly with the two atomic species, $\upp$ and $\dnn$, that make up the superfluid (i.e. $\upp$ and $\dnn$ do not become localized). The magnetic character of the impurity originates in the different interactions strengths between the impurity atom and $\upp$ and $\dnn$ atoms, which we describe by a pair of effective scattering lengths $a_\uparrow$ and $a_\downarrow$. 

The main input into our theory of impurity localized states is the description of the atom scattering on a localized impurity. (1) We begin by showing that, under rather general conditions, we can describe the interaction between a localized impurity and the free atoms via an effective s-wave scattering length. (2) As pointed out by Shiba, due to the sharpness of the BCS density of states, the magnetic impurity always results in the formation of a pair of localized bound states, called the Shiba states. Indeed, we find that as long as $a_\uparrow \neq a_\downarrow$, the impurity atoms always induce a pair of bound states inside the superconducting gap. (3) Interestingly, we find that the Shiba state is not related to the under-sea bound state. By the under-sea bound state we mean the natural extension to the case of a filled Fermi sea of the Feshbach bound state formed between a fermion and a localized impurity  in the absence of a Fermi sea when the effective scattering length is positive. Indeed, if the Feshbach bound state exists, it becomes the under-sea bound state when the Fermi sea is filled, remaining completely separate of the Shiba state. (4) We show that both the under-sea bound states and the Shiba bound states can be resolved via RF spectroscopy.

This paper is organized as follows: In section~\ref{sec:scatteringLength} we relate the bare scattering length between a pair of atoms in vacuum to the effective scattering length when one of the atoms is localized by a parabolic confining potential. In section~\ref{sec:boundStates} we use the effective scattering lengths to find the under-sea as well as the Shiba bound states of the magnetic impurity. Next, we describe RF spectroscopy of the ultracold atom system with bound states in section~\ref{sec:spectroscopy}. We discuss possible experimental realizations and atom species that could be used in section~\ref{sec:experiment}, discuss the outlook in section~\ref{sec:outlook}, and draw conclusions in section~\ref{sec:conclusions}.

\section{Efective Scattering Length}
\label{sec:scatteringLength}
The goal of this section is to show that the scattering of a fermionic atom off of a confined impurity can, under reasonable conditions, be described by a single quantity -- the effective scattering length.  The problem of scattering on a confined impurity was previously studied in Ref.~\cite{Castin}, here we review the basic arguments and summarize the results.

We begin by assuming that the impurity-fermion scattering in vacuum can indeed be defined by a single scattering length for the s-wave scattering process. This condition means that the effective range $r_0$ of the impurity-fermion interaction potential is much smaller than the typical fermion wavelength $1/k_F$, and thus we can treat $r_0$ as being essentially zero.  Since we want the fermion-impurity interaction to be tunable, we shall be primarily interested in operating in the vicinity of a wide Feshbach resonance (i.e. a resonance that meets the condition $r_0 \ll 1/k_F$). If the effective range condition is not satisfied for the case of a free impurity (e.g. for the case of a narrow Feshbach resonance), it will not be satisfied for the case of a localized impurity, necessitating a more complicated description of the effective scattering process. Although, we do not treat the more complicated case in the present paper,  we expect that the qualitative features, including the Shiba bound states, of a system with a narrow impurity resonance will be similar to those of a system with a wide resonance.

In the problem with a confined impurity we have two important energy scales: the typical kinetic energy of a scattering fermion, which in our case is set by the Fermi energy scale $\epsilon_F$, and the level spacing of the impurity atom which we label $\hbar \omega_i$.  We begin by pointing out that the scattering is elastic in the regime $\epsilon_F \ll \hbar \omega_i$. Further, in order for the scattering to be dominated by s-wave channel, we demand that the ground state wavefunction of the impurity must have a length scale $\sqrt{\hbar/m_i\omega_i}$ that is much smaller than the wavelength of the scattering particle $\hbar/\sqrt{2m_\alpha\epsilon_F}$ (here $m_i$ stands for the mass of the impurity and $m_\alpha$ for the mass of the scattering fermion).  The two conditions are identical, up to a ratio of the masses. That is we demand that $\text{max}(1,m_\alpha/m_i) \epsilon_F \ll \hbar \omega_i$.

Having derived the conditions for s-wave scattering we can write the resulting T-matrix for the scattering atom in the form 
\begin{align}
T(\omega)=\frac{1}{\frac{m_\alpha}{2\pi}\left(\frac{1}{a_\alpha}+i\sqrt{2m \omega}\right)}.
\end{align}
It is important to point out that since the impurity is localized the T-matrix features the fermion mass as opposed to the reduced mass $\mu=\left(m_i^{-1}+m_\alpha^{-1}\right)^{-1}$ and an effective scattering length $a_\alpha$ instead of the vacuum scattering $a_{0,\alpha}$. In order to relate the effective scattering length to the vacuum scattering length, we must solve the scattering problem. In general the scattering problem is complicated, and requires a numerical solution. In appendix~\ref{app:Scattering}, we state the scattering problem and derive an analytic solution for the special case of weak impurity-fermion interactions using a Born-Oppenheimer type approximation.

\section{Impurity Bound States}
\label{sec:boundStates}
In this section, we study the conditions for the existence of impurity bound states both in the normal (single component, non-interacting Fermi gas) and in the superconducting case. Our strategy is to obtain the T-matrix for scattering off of an isolated impurity in the presence of the Fermi-sea. Having the T-matrix, we can find the energies of the bound states from its poles. Further, we can also find the spectral function of the fermions, which we shall use in the next section to compute the RF spectra. 

\begin{figure}
\includegraphics[width=8cm]{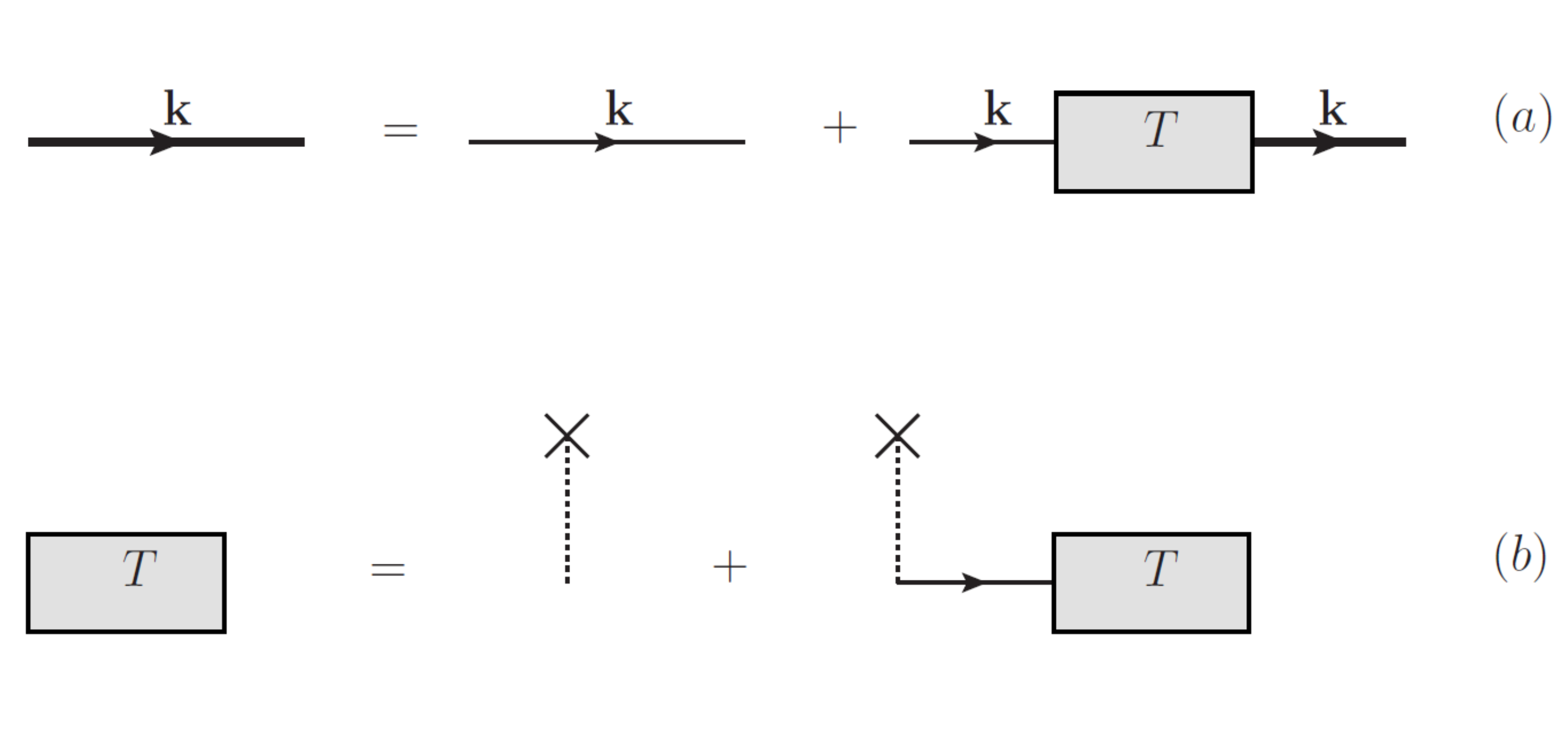}
\caption{(a)  diagrammatic representation of equation \eqref{GG0T}  and (b) of equation \eqref{Lippmann-Schwinger}. Thin lines represent the clean (unperturbed) Green functions ($G^0(\kk, \omega)$), the thick lines the impurity-perturbed Green function ($G_\textbf{k}$), and the dashed line the interaction of these fermions with the impurity ($V$)}
	\label{fig:feynman}
\end{figure}

In general, we can express the effect of the impurities on the Green function of the clean system $G^0(\kk,\omega)$ via an expansion in the impurity density~\cite{Mahan}
\begin{align} 
G(\kk,\omega)&=G^0(\kk,\omega)+n_i G^0(\kk,\omega)T(\omega)G^0(\kk,\omega) \label{GG0T}  \\
	& \hspace{5cm} + O(n_i^2)  \nonumber
\end{align}
where $G(\kk,\omega)$ is the Green function of the dirty system, $T(\omega)$ is the T-matrix, and $n_i$ is the impurity density. In this paper we shall always work in the dilute impurity limit, and thus drop terms of order $O(n_i^2)$ and higher. The resulting equation is represented diagrammatically in Fig.~\ref{fig:feynman}a. $T(\omega)$ is obtained from the Lippmann-Schwinger equation
\begin{equation} 
T(\omega)=V+V \mathcal{G}^0(\omega) T(\omega), \label{Lippmann-Schwinger} 
\end{equation}
which relates the T-matrix to the impurity-fermion interaction potential $V$ and the momentum integrated Green function of the clean system
\begin{equation}
\mathcal{G}^0(\omega)=\int \frac{d^3 \kk}{(2\pi)^3} G^0(\kk,\omega).
\end{equation}
The Lippmann-Schwinger equation is illustrated diagrammatically in Fig.~\ref{fig:feynman}b; it can be formally solved for the T-matrix by inversion
\begin{equation} 
T^{-1}(\omega)=V^{-1}-\mathcal{G}^0(\omega). \label{eq:TMatrix}
\end{equation}

Having specified the Lippmann-Schwinger equation, we first apply it to the case of an impurity in a one component non-interacting Fermi gas. This trivial case serves as an exercise that demonstrates (1) regularization of point contact interactions, (2) properties of the T-matrix, and (3) relation between Feshbach molecules and under sea states. Having learned how to use the Lippmann-Schwinger equation in this context, we apply it to the T-matrix of the BCS state.

\subsection{Impurity in a one component non-interacting Fermi gas}

We first consider a fixed impurity interacting with a one component Fermi sea, the interaction being described by the scattering length $a$. Since the Fermi sea is non-interacting and the impurity is static, we can proceed simply by finding the one-particle eigenstates in the vicinity of the impurity potential, and filling them up to the Fermi energy. For negative $a$, all eigenstates are part of the continuum, and there are no localized bound states on the impurity. As $a$ becomes positive, a single state, the Feshbach molecular state, with energy $-1/(2 m a^2)$ peels off the continuum and becomes localized by the impurity (henceforth, the mass of the impurity no longer features and therefore we will use $m$ for the mass of the fermion). When we fill the Fermi sea, the Feshbach molecular state appears as an under-sea bound state.

In this subsection, we show how to recover this simple picture in the T-matrix language. In the absence of impurity, the fermions are described by the following Green function~\cite{Mahan}
\begin{equation} 
G(\kk, \omega)=\frac{1}{\omega-\xi_\kk+i0^+ \text{sgn}(\omega)},  \label{eq:1GF}
\end{equation}
where $\xi_\kk \equiv \epsilon_\kk-\epsilon_F \equiv \frac{\hbar^2\kk^2}{2m}-\epsilon_F$ and $\epsilon_F$ is the Fermi energy. In order to cancel the divergence of the integral of the Green function in the Lippmann-Schwinger equation we must use a renormalized interaction potential~\cite{KetterleZwierleinNotes, Marini}
\begin{equation}
\frac{1}{V}=\frac{2m}{4\pi\hbar^2 a}-\frac{2m}{\hbar^2}\int \frac{d^3\kk}{(2\pi)^3}\frac{1}{\kk^2}.
\end{equation}
As described in Sec.~\ref{sec:scatteringLength}, because the impurity atom is confined in the expression for the interaction potential we must use the fermion mass $m$ and the effective scattering length $a$ instead of the reduced mass and the vacuum scattering length. The momentum integrated Green function $\mathcal{G}^0$, that enters the Lippmann-Schwinger equation, can be obtained via contour integration
\begin{equation} 
\mathcal{G}^0(\omega)=\int \frac{d^3\kk}{(2\pi)^3} \frac{1}{\epsilon_\kk}-i\frac{(2m)^{3/2}}{4\pi}\sqrt{\omega+\epsilon_F}.
\end{equation}
The divergence in $\mathcal{G}^0(\omega)$ is perfectly canceled by the renormalized interaction to yield the T-matrix
\begin{equation} 
T(\omega)=\frac{1}{\frac{m}{2\pi}\left(\frac{1}{a}+i\sqrt{2m(\omega+\epsilon_F)}\right)}. \label{eq:1T}
\end{equation}
Unsurprisingly, the T-matrix has the same form as the vacuum T-matrix, but with frequency shifted by $\epsilon_F$. This reflects the fact that energies must be measured with respect to the Fermi energy.  
The bound states of the system introduced by the presence of the impurity are defined by the poles of the T-matrix. We thus find that a bound state exists only for positive values of the scattering length, with an energy
\begin{equation} 
\omega_b=-\epsilon_F-\frac{1}{2ma^2}.
\end{equation}

\subsection{Impurity in BCS state}
In this subsection, we generalize the results of the previous subsection to the case of a localized impurity atom immersed in an ultracold BCS gas. We shall describe the BCS state at the mean-field level. Since BCS quasi-particles involve mixing particles and holes, it is convenient to use Nambu's 4-dimensional spinor basis~\cite{Nambu, Shiba, Ambegaokar}
\begin{equation} 
\Psi_\textbf{k}= \left(\begin{array}{c} c_{\textbf{k}\uparrow} \\ c_{\textbf{k}\downarrow} \\ c_{\textbf{-k}\uparrow}^\dag \\ c_{\textbf{-k}\downarrow}^\dag \end{array}\right)
\end{equation}
In this formalism, the BCS Hamiltonian becomes
\begin{equation} 
H_\text{BCS}=\left(\begin{array}{cccc} 
\xi_\kk & 0 & 0 & -\Delta \\ 
0 & \xi_\kk & \Delta & 0 \\ 
0 & \Delta & -\xi_\kk & 0 \\ 
-\Delta & 0 & 0 & -\xi_\kk
\end{array}\right),
\end{equation}
where $\Delta$ is the BCS order parameter. The BCS Green function of the clean system is
\begin{align}
&G^0(\kk,\omega) = \frac{1}{\omega-\xi_\textbf{k}\rho_3-\Delta\sigma_2\rho_2} = 
\frac{\omega + \xi_\textbf{k}\rho_3 + \Delta\sigma_2\rho_2}{\omega^2 - \xi_\textbf{k}^2-\Delta^2} \nonumber \\ 
&\quad\equiv  \frac{1}{\omega^2-\xi_\textbf{k}^2-\Delta^2} \left(\begin{array}{cccc} \omega+\xi_\textbf{k} & 0 & 0 & -\Delta \\ 
0 & \omega+\xi_\textbf{k} & \Delta & 0 \\
0& \Delta &\omega-\xi_\textbf{k} & 0 \\
-\Delta & 0 & 0 & \omega-\xi_\textbf{k}
\end{array}\right). \label{eq:GBCS}
\end{align}
Here, $\{\sigma_1,\sigma_2,\sigma_3\}$ and $\{\rho_1,\rho_2,\rho_3\}$ are two sets of Pauli matrices, the first one operating on the spin space and the second on the particle-hole space. 

The interaction potentials between each of the two species that make up the BCS state and the impurity atom have the same form as the interaction potential in the single component case
\begin{equation}
\frac{1}{V_{\uparrow (\downarrow)}}=\frac{2m}{4\pi\hbar^2 a_{\uparrow (\downarrow)}}-\frac{2m}{\hbar^2}\int \frac{d^3\kk}{(2\pi)^3}\frac{1}{\kk^2}.
\end{equation}
Here $a_\uparrow$ corresponds to the effective scattering length between a $\upp$ atom and the localized impurity, while $a_\downarrow$ between a $\dnn$ atom and the impurity. In Nambu basis, the interaction potential becomes
\begin{equation} 
V=\left(\begin{array}{cc} 
\frac{1}{V_1} &0\\
0& \frac{1}{V_2}
\end{array}\right)\otimes\rho_3=\left(\begin{array}{cccc} \frac{1}{V_1} &0&0&0 \\ 0 &\frac{1}{V_2} & 0&0 \\ 0& 0&-\frac{1}{V_1}&0\\0 &0 & 0&-\frac{1}{V_2} 
\end{array}\right). \end{equation}

Substituting $G^0(\kk,\omega)$ and $V$ into the T-matrix equation \eqref{eq:TMatrix} we see that the four-dimensional Nambu space reduces into a pair of two-dimensional subspaces that can treated separately: the `outer' (or `first') subspace acts on the first and fourth Nambu components, whereas the `inner' (or `second') subspace acts on the second and third components. From this point, we will limit ourselves to one of them, say the first one.

To use Lippmann-Schwinger equation in order to yield the T-matrix, our first step is the calculation of $\mathcal{G}^0$, which can be written as the sum of a regular part $\mathcal{G}^0_r(\omega)$ and a diverging part, as
\begin{equation} \mathcal{G}^0(\omega)=\mathcal{G}^0_r(\omega)-\int \frac{d^3\kk}{(2\pi)^3} \frac{1}{\epsilon_\kk} \rho_3.
\end{equation}
From this definition, the regular part is 
\begin{widetext}
\begin{align} 
\mathcal{G}^0_r(\omega) &= \int \frac{d^3\kk}{(2\pi)^3} 
\left[
\frac{1}{\omega^2-\xi_\textbf{k}^2-\Delta^2} 
\left(\begin{array}{cc} \omega & -\Delta \\ -\Delta & \omega \end{array}\right)
+\left(\frac{\xi_\textbf{k}}{\omega^2-\xi_\textbf{k}^2-\Delta^2}+\frac{1}{\epsilon_\textbf{k}}\right)
\left(\begin{array}{cc} 1 & 0 \\ 0 & -1 \end{array}\right)
\right].\end{align}
\end{widetext}
$\mathcal{G}^0_r(\omega)$ can be expressed as a function of two integrals 
\begin{align} 
I_1(\omega)&=\int_0^{\infty}\frac{\kappa^2 d\kappa}{\omega^2-\Delta^2-(\kappa^2-\epsilon_F)^2}, \\
I_2(\omega)&=\int_0^\infty\frac{d\kappa}{\omega^2-\Delta^2-(\kappa^2-\epsilon_F)^2},
\end{align} 
where we have used the notation $\kappa = \kk/\sqrt{2m}$. Both integrals can be evaluated using contour integration, to give
\begin{widetext}
\begin{align} I_1(\omega)&=\frac{1}{4\pi\sqrt{\Delta^2-\omega^2}}\left[\sqrt{\epsilon_F+i\sqrt{\Delta^2-\omega^2}}+\sqrt{\epsilon_F-i\sqrt{\Delta^2-\omega^2}}\right] \\
I_2(\omega)&=\frac{1}{4\pi\sqrt{\Delta^2-\omega^2}}\frac{\sqrt{\epsilon_F+i\sqrt{\Delta^2-\omega^2}}+\sqrt{\epsilon_F-i\sqrt{\Delta^2-\omega^2}}}{\sqrt{\epsilon_F^2+\Delta^2-\omega^2}} \end{align}
\end{widetext}
Using the fact that 
\begin{align} 
&\sqrt{\epsilon_F+i\sqrt{\Delta^2-\omega^2}}+\sqrt{\epsilon_F-i\sqrt{\Delta^2-\omega^2}} \nonumber \\
&\quad\quad\quad= 2\left(\frac{\epsilon_F+\sqrt{\epsilon_F^2+\Delta^2-\omega^2}}{2}\right)^{1/2}
\end{align}
we find
\begin{align} 
\mathcal{G}^0_r(\omega)& = - i \frac{m^{3/2} (\epsilon_F+\Xi)^{1/2}}{2 \pi \sqrt{\omega^2-\Delta^2}}
\text{sgn}(\Re(\omega)\Im(\omega))
\nonumber \\
& \quad\quad\quad\quad \left(\begin{array}{cc} \omega+(\epsilon_F-\Xi) & -\Delta \\ 
-\Delta & \omega-(\epsilon_F-\Xi)
\end{array}\right), 
\end{align} 
where $\Xi=\sqrt{\epsilon_F^2+\Delta^2-\omega^2}$, and the $\text{sgn}$ function ensures that we take the correct branch of the square roots. Once more, the two diverging integrals in $\mathcal{G}^0$ and $V^{-1}$ cancel, and \eqref{eq:TMatrix} yields
\begin{align} 
T^{-1}(\omega) = &\left(
\begin{array}{cc} 
\frac{m}{2 \pi a_\uparrow} & 0\\ 
0 & -\frac{m}{2 \pi a_\downarrow}
\end{array}\right) - \mathcal{G}^0_r(\omega)  
\label{eq:fullTMatrix}
\end{align}
Having solved the Lippmann-Schwinger equation, we can look at the properties of the resulting T-matrix. In particular, we want to consider two regimes: bound states inside the gap and bound states outside the gap.

\subsubsection*{Under-sea states}
\label{sec:undersea}
Aiming to recover the Feshbach molecule-like bound state that we found to exist under the Fermi sea in the case of a one component gas, we make the approximation that $\Delta\simeq 0$. The bound state must correspond to a frequency $\omega=\omega_b+i 0^+$, where $\omega_b\leq -\epsilon_F$. Within this approximation,
 \begin{align} 
T^{-1}(\omega) \approx 
&\left(
\begin{array}{cc} 
\frac{m}{2 \pi a_\uparrow} & 0\\ 
0 & -\frac{m}{2 \pi a_\downarrow}
\end{array}\right) \\
&-\frac{\sqrt{2}\,m^{3/2}}{2 \pi}
\left(\begin{array}{cc} 
\sqrt{-\epsilon_F-\omega} & 0
\\ 0 & - i \sqrt{\epsilon_F-\omega}
\end{array}\right) \label{eq:Tundersea}
\end{align}
We find that the T-matrix only has a pole (i.e. $\det T^{-1}(\omega)=0$) when the effective scattering length $a_\uparrow$ is positive. The frequency of the pole is
\begin{equation}
\omega_b=-\epsilon_F-\frac{1}{2ma_\uparrow^2}
\end{equation} 
By looking at the complimentary $2\times2$ Nambu subspace, we find that another bound state exists for positive values of $a_\downarrow$ with frequency $\omega_b=-\epsilon_F-1/2ma_\downarrow^2$.

If we relax the approximation $\Delta\simeq 0$, we find that the under-sea state only becomes a sharp bound state in the limit $\omega_b\rightarrow -\infty$. If the binding energy is not very large, then the under-sea bound state can serve as a Kondo impurity. However, detailed analysis of this possibility is beyond the scope of the present article. 

\subsubsection*{Shiba states}
\label{sec:Shiba}
\begin{figure}
\includegraphics[width=7cm]{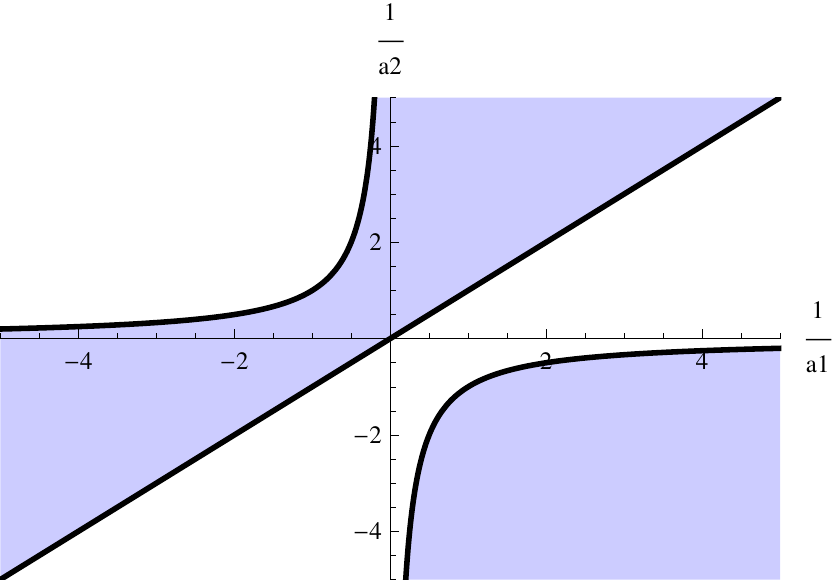}
\caption{Representation of the sign of the solutions given by the first and second subspaces as a function of $1/k_F a_\uparrow$ and $1/k_F a_\downarrow$. In the blue colored domain the solution given by the first subspace is negative, while in the white colored domain the one given by the second subspace is negative. }
\label{fig:a1a2plane}
\end{figure}

\begin{figure}
\includegraphics[width=8cm]{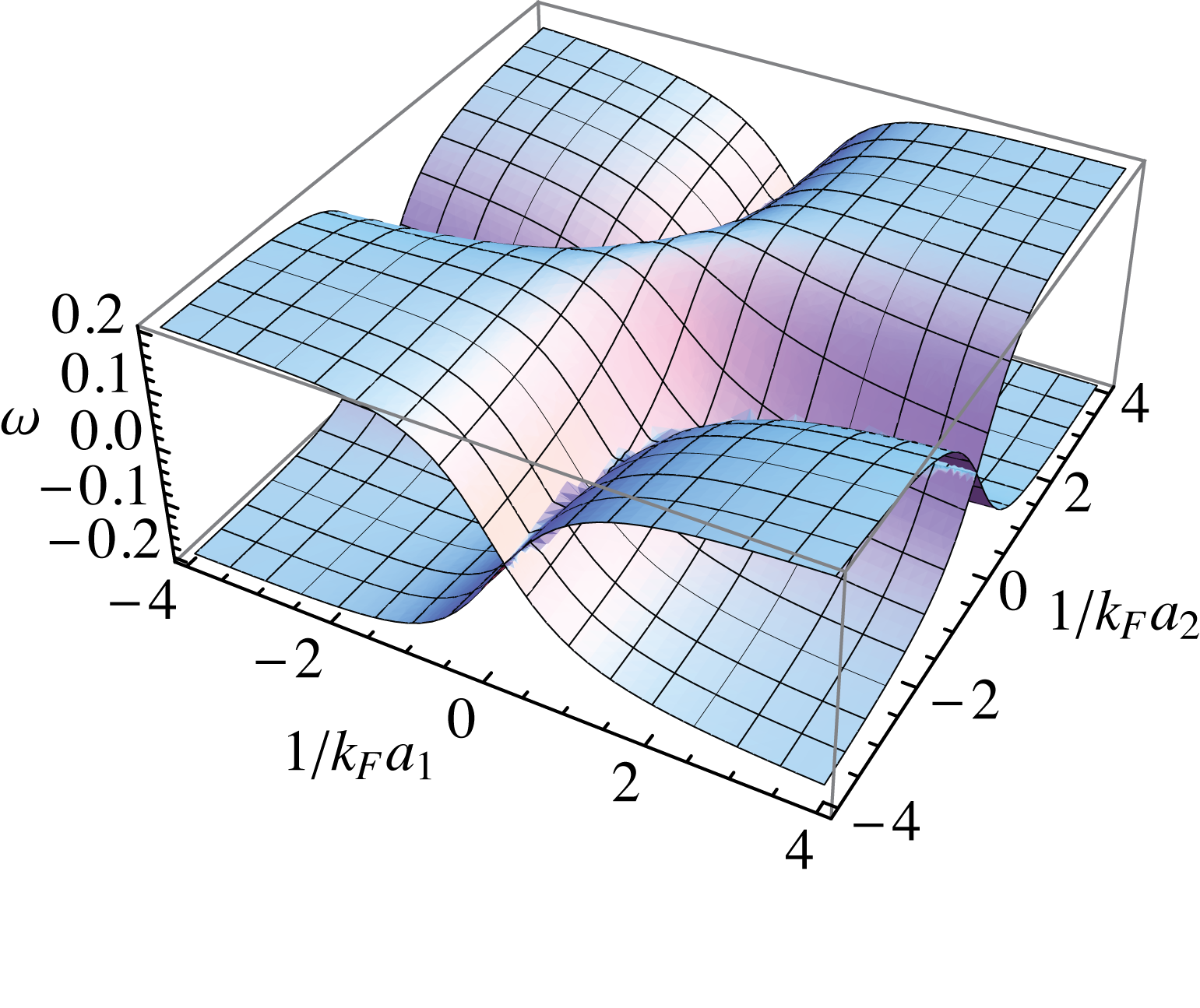}
\caption{Energies of the two in-gap (Shiba) bound states as a function of $1/k_F a_\uparrow$ and $1/k_F a_\downarrow$ (here we use the approximation of \eqref{eq:Freq}, we took $\Delta=0.2 \epsilon_F$, and $\omega$ is measured in units of $\epsilon_F$)}
\label{fig:grapheomega}
\end{figure}

We now turn to the in-gap bound states predicted by Shiba, that is, $|\omega|<\Delta$. For weakly enough interacting BCS gases, we can make the approximation $|\omega|<\Delta\ll\epsilon_F$. Within this approximation,
\begin{align} 
T^{-1}(\omega) \approx &
\left(\begin{array}{cc} 
\frac{m}{2 \pi a_\uparrow} & 0\\ 
0 & -\frac{m}{2 \pi a_\downarrow}
\end{array}\right) \nonumber
\\
&\quad\quad\quad +\frac{m^{3/2} \sqrt{2\epsilon_F}}{2 \pi \sqrt{\Delta^2-\omega^2}}
\left(\begin{array}{cc} 
\omega & -\Delta\\ 
-\Delta & \omega
\end{array}\right) \label{Tintragap}
\end{align}
The form of the T-matrix in the complimentary Nambu subspace can be obtained from this one by making the substitutions $\Delta\to-\Delta$, $a_\uparrow \leftrightarrow a_\downarrow$. The poles of the T-matrix are defined by the equation 
\begin{equation}
\frac{\omega}{\sqrt{\Delta^2-\omega^2}} = \pm  \frac{1+k_F a_\uparrow k_F \, a_\downarrow}{k_F a_\downarrow-k_F a_\uparrow}, \label{eq:Freq}
\end{equation}
where the $+$ sign corresponds to the first Nambu subspace and the $-$ sign to the second Nambu subspace.  From equation \eqref{eq:Freq}, we see that as long as $a_\uparrow \neq a_\downarrow$ there is exactly one pole of the T-matrix in each of the two subspaces. The two poles have opposite frequencies and correspond to the two Shiba states. We can interpret the negative frequency pole as a bound state for the quasi-particle of the gas, and the positive frequency solution as a bound quasi-hole. In Fig.~\ref{fig:a1a2plane}, we split the $\left\{1/k_F a_\uparrow,\,1/k_F a_\downarrow\right\}$ plane into two domains: the blue domain corresponds to negative pole being in the first subspace, and the white domain to the negative pole in the second subspace. The corresponding frequencies of the two Shiba states are plotted as a function of $1/k_F a_\uparrow$ and $1/k_F a_\downarrow$ in Fig.~\ref{fig:grapheomega}.

From \eqref{eq:Freq} and Fig.~\ref{fig:grapheomega}, we see that the bound states are located inside the gap only for nonzero values of $a_\uparrow-a_\downarrow$. This fact can be quite straightforwardly interpreted: the interaction between the Cooper pairs and the impurity can be analyzed as the sum of a `magnetic' term proportional to $a_\uparrow-a_\downarrow$ and a non-magnetic term proportional to $a_\uparrow+a_\downarrow$. The impurity can break Cooper pairs and give rise to in-gap states only when the magnetic term is finite. We note that when the non-magnetic term becomes zero, we recover the formula established by Shiba for a spin impurity in an electronic superconductor. 

\begin{figure}
\includegraphics[width=7cm]{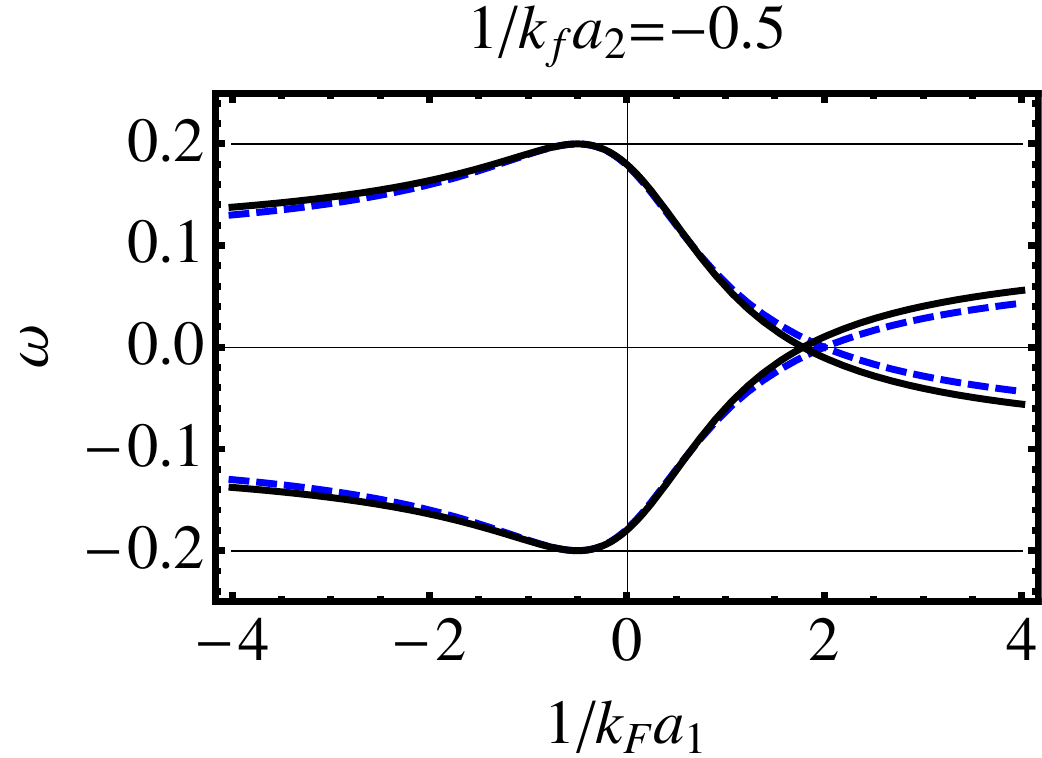}
\caption{Comparison between the approximate analytical solution (blue dashed curve) and the exact numerical solution (black curve) for the frequency (in units of $\epsilon_F$) of the in-gap bound state (of the first Nambu subspace) as a function of $1/k_F a_\uparrow$, with $\Delta=0.2 \epsilon_F$ and $1/k_F a_\downarrow=-0.5$}
\label{fig:error}
\end{figure}

Finally, we comment on the approximation that went into \eqref{eq:Freq}. In figure \ref{fig:error} we compare the frequency of the Shiba state (of the first Nambu subspace) calculated using \eqref{eq:Freq} and numerical solution of \eqref{eq:fullTMatrix}. We see excellent agreement between the approximate and exact answers, which persists to surprisingly large values of gap, $\Delta \lesssim 0.5 \epsilon_F$.

\subsubsection*{Discussion of bound states}
We underline that the Shiba and under-sea bound states are not related to each other. For example if both scattering lengths are negative, but unequal, then the two Shiba states are still present while the under-sea states are not. On the other hand for two positive and unequal scattering lengths there is a pair of under-sea bound states in addition to the two Shiba states. Finally if one scattering length is positive and the other is negative then there are again two Shiba states but only one under-sea state.

\begin{figure}
\includegraphics[scale=0.5]{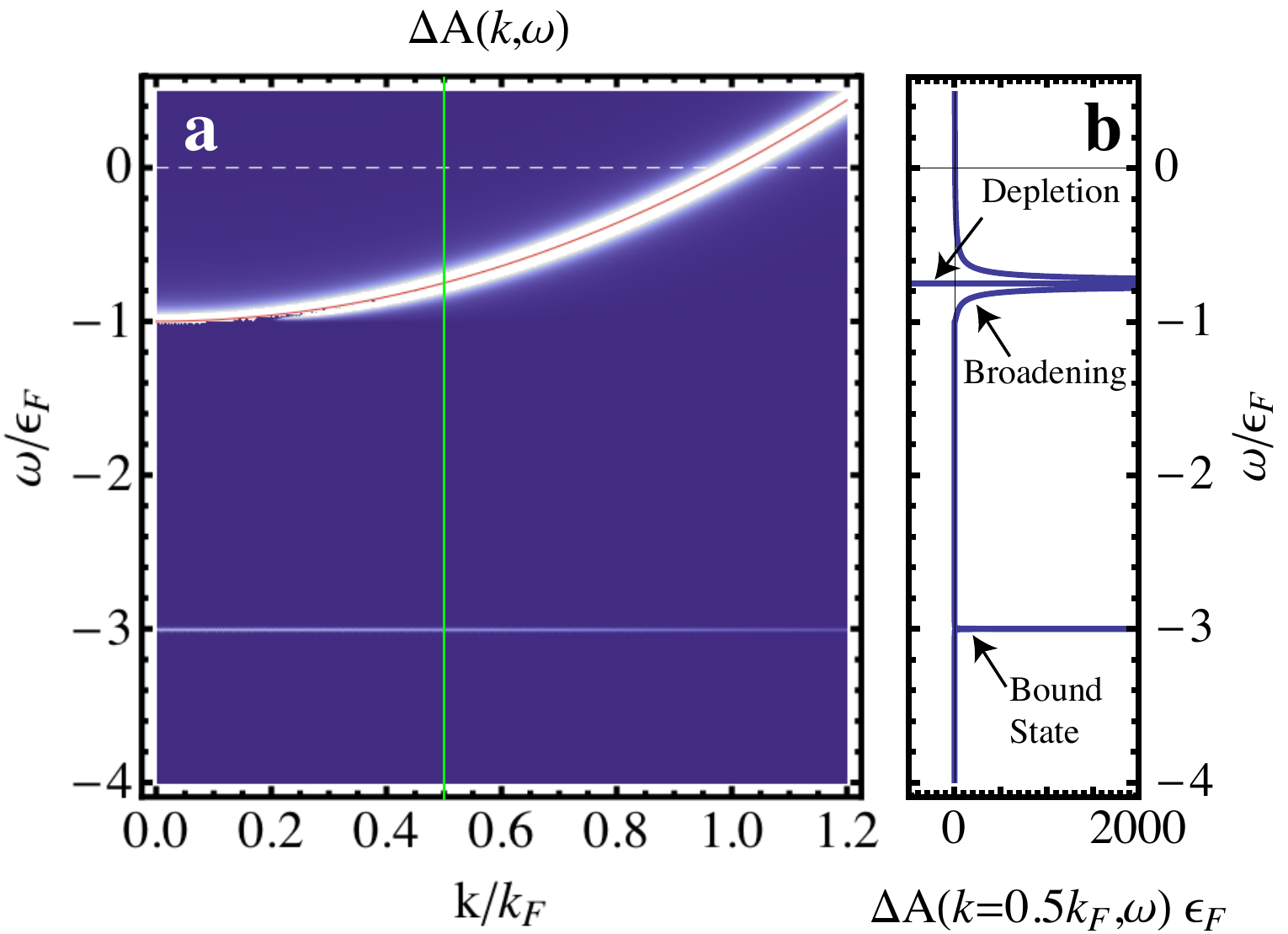} 
\caption{Impurity induced correction to the spectral function $\Delta A(\kk, \omega)$ of the one component Fermi gas for the case $a k_F=0.5$ (white -- increase of spectral weight, blue -- no change, red -- decrease). (a) $\Delta A(\kk, \omega)$ as a function of momentum and frequency. The dashed white line indicates the position of the Fermi energy.  $\Delta A(\kk, \omega)$ shows a depletion of spectral weight along the clean dispersion line $k^2/2m-\epsilon_F$ (indicated by the red line), an under-sea bound state at $\omega=-3 \epsilon_F$, and excess spectral weight in the vicinity of the continuum band which corresponds to the impurity induced broadening. (b)  $\Delta A(\kk, \omega)$ as a function of frequency only with momentum fixed at $k=0.5k_F$ [slice is indicated by the green line in (a)]. The spectral function can be decomposed into three (labeled) features: (1a) a $\delta$-function corresponding to the depletion of spectral weight along the clean dispersion line; (1b) part of the depleted weight is transferred into the vicinity of the clean dispersion line resulting in its broadening; (2) the remaining weight is transferred to a $\delta$-function corresponding to the under sea bound state. We note that although the part labeled ``Broadening" is divergent in the impurity density expansion, its frequency integral remains finite, and the spectral function fulfills the frequency sum rule.
}
\label{fig:A1I}
\end{figure}

\begin{figure}
\includegraphics[width=8cm]{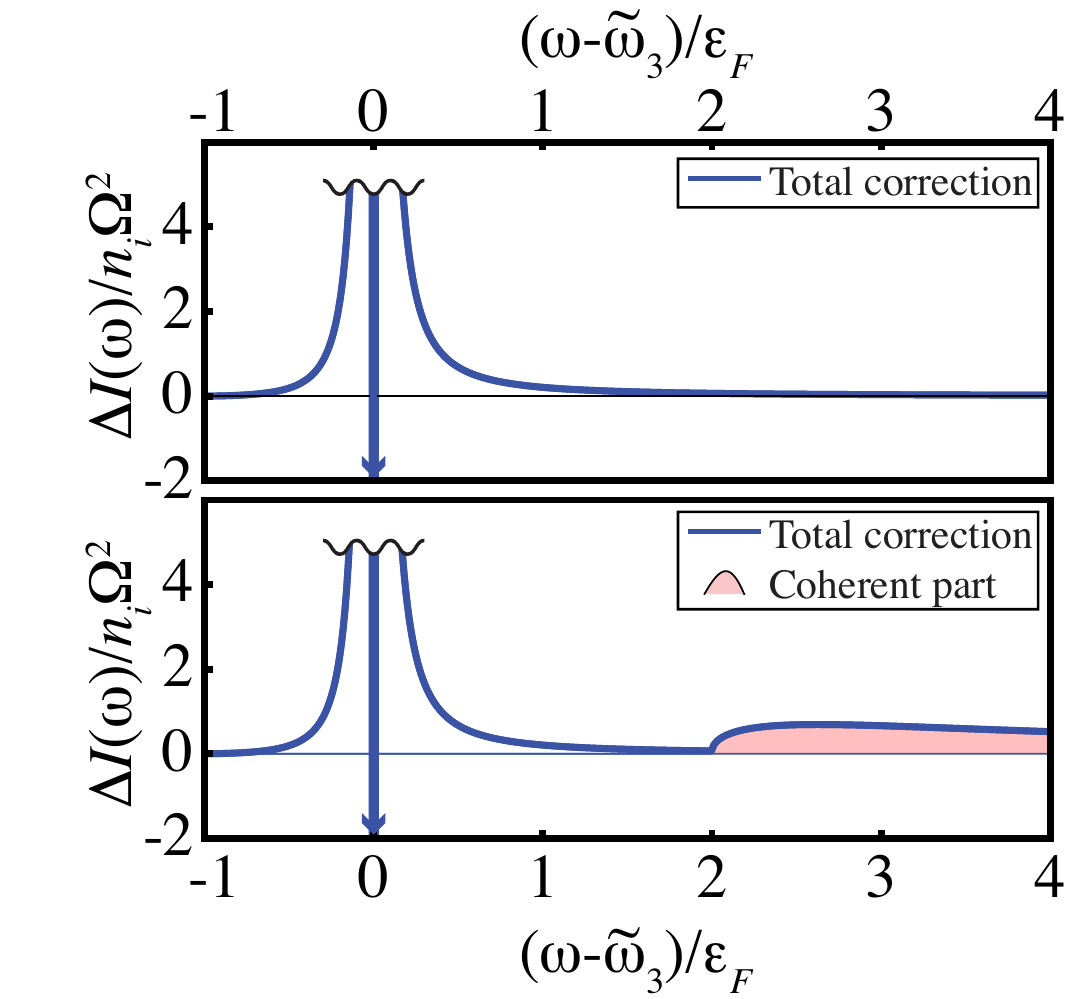}
\caption{Correction to the RF transition rate obtained for the one component gas due to the presence of impurities as a function of the drive frequency $\omega$, with $k_F a=-0.5$ (top) and $k_F a=0.5$ (bottom). The RF spectrum for the clean case is sharply peaked at $\omega-\tilde{\omega}_3 \sim 0$ with the width set by either trap properties and temperature. The impurities have two main effects: (1) Since momentum is no longer a good quantum number, the impurities broaden the sharp absorbtion peak at $\omega-\tilde{\omega}_3 \sim 0$. This broadening is composed of the depletion of the $\delta$-function indicated by the blue arrow together with population of nearby-in-frequency states. (2) If there is a bound state, it induces an edge in the spectrum of transferred atoms followed by a broad feature indicated in pink. The broadening correction cannot be accurately captured in an expansion in impurity density. In fact at first order in impurity density we find that the correction is divergent but integrable. Therefore, in the figure we cut it off with a wavy line. While feature (1) is present independently of the sign of the scattering length, feature (2) which corresponds to the coherent part of the transition rate correction (i.e. the bound state induced part) is present only for positive scattering length. }
\label{fig:graphnormal}
\end{figure}

\section{RF Spectroscopy}
\label{sec:spectroscopy}
We suggest that radio-frequency spectroscopy could be a good experimental probe for reading out properties of the Shiba as well as under-sea bound states. Basic tools for understanding RF spectroscopy are given in \cite{KetterleZwierleinNotes}. RF spectroscopy works by converting $\upp$ (or equivalently $\dnn$) atoms to a third hyperfine state labeled $|3\rangle$ by irradiating the system with photons of frequency $\omega_\text{RF}$ that bridges the energy difference between  $\upp$ and $|3\rangle$ states. The bound states show up as edges in the spectra of transferred atoms when $\omega_\text{RF}$ matches the bound state energy.

In the following, we begin by reviewing the Fermi golden rule formula, in terms of $\upp$ Green function, for the $\upp \to |3\rangle$ transition rate as a function of $\omega_\text{RF}$. Next, we apply the formula first to the case of one component gas and second to the BCS case.

\subsection{General formula for the RF transition rate}
We assume that the Hamiltonian of the system, subject to RF drive, may be written in the form
\begin{equation}
H=H_{\text{gas},\text{impurity}} + H_3 + H_\text{RF},
\end{equation}
where $H_{\text{gas},\text{impurity}}$ describes the fermion gas and the impurity, $H_3$ describes the Fermions in the $|3\rangle$ hyperfine state, and $H_\text{RF}$ describes the action of the RF radiation. In writing $H$ in this form, we make the standard assumption that fermions in the $\upp$ and $\dnn$ hyperfine states do not interact with fermions in the $|3\rangle$ hyperfine state except through the action of $H_\text{RF}$. Our goal is to calculate the RF current (i.e. the transfer rate of atoms from state $\upp$ to state $|3\rangle$) that is induced by $H_\text{RF}$, which we do in second order perturbation theory (Fermi golden rule).

The RF drive can be described by the Hamiltonian
\begin{equation} 
H_\text{RF}=\Omega_{RF}\int \frac{d^3\kk}{(2\pi)^3}(e^{-i \omega_\text{RF} t}c_{3,\kk}^\dag c_{\uparrow, \kk}+e^{i \omega_\text{RF} t}c_{\uparrow,\kk}^\dag c_{3,\kk}), 
\end{equation}
where $\Omega_\text{RF}$ and $\omega_\text{RF}$ are the intensity and frequency of the RF drive; $c_{3,\kk}^\dag$ ($c_{3,\kk}$) and $c_{\uparrow,\kk}^\dag$ ($c_{\uparrow,\kk}$) are the creation (annihilation) operators for fermions in the $\upp$ and $|3\rangle$ hyperfine states. Since RF photons have a very small momentum (large wavelength) we neglect the momentum imparted on the atoms by the photons. Atoms in $|3\rangle$  hyperfine state are treated as free fermions and are described by the Hamiltonian
\begin{align}
H_3=\int \frac{d^3\kk}{(2 \pi)^3} (\omega_3+\epsilon_\kk) c_{3,\kk}^\dag c_{3,\kk},
\end{align}
where $\omega_3$ is the splitting between the $\upp$ and $|3\rangle$ states in vacuum. The corresponding (Matsubara) Green function for  $|3\rangle$ fermions is
\begin{align}
G_3(\kk, i \omega_n)=\frac{1}{i \omega_n-(\epsilon_\textbf{k}+\omega_3)}. \label{eq:3GF}
\end{align}

The Golden rule formula states that current from $\upp$ to $|3\rangle$ is~\cite{Mahan}
\begin{align}
I(\omega_\text{RF})=2 \Omega^2 \Im \left[ \mathcal{D}(i \omega_n\to \omega_\text{RF} + i 0^+) \right], \label{eq:I}
\end{align}
where
\begin{equation} 
\mathcal{D}(i \omega_n)=\int\frac{d^3\kk}{(2\pi)^3} \frac{1}{\beta}\sum_{i \omega_1}G_\uparrow(\kk, i \omega_1)G_3(\kk, i \omega_1 + i \omega_n),\label{eq:D}
\end{equation}
and $\omega_1$ and $\omega_n$ are fermionic and bosonic Matsubara frequencies, respectively. Our Golden rule formula gives the transition rate per unit volume. To obtain the transition rate per particle, we must divide $I(\omega_\text{RF})$ by density [we shall use units where the density is set to $k_F^3/(6 \pi^2) =\sqrt{2}/(3 \pi^2)$]. We restate the golden rule formula in the more familiar real time version
\begin{align}
I(\omega_\text{RF})=\Omega^2  \int \frac{d^3\kk}{(2\pi)^3} \frac{d\epsilon}{2 \pi}   A_\uparrow(\kk,\epsilon) A_3(\kk,\epsilon+\omega_\text{RF}) n_F(\epsilon), \label{eq:IA}
\end{align}
where $A_\sigma(\kk,\omega)=-2 \Im G_\sigma(\kk,\omega+i0^+)$ are the spectral functions for $\sigma=\{\uparrow,3\}$ fermions, $n_F(\epsilon)$ is the Fermi function for the $\uparrow$ fermions, and we have assumed that the $3$ band is empty. Using the fact that the $|3\rangle$ state is non-interacting, we can simplify this expression
\begin{align} 
I(\omega_\text{RF})=&\Omega^2 \int \frac{d^3\kk}{(2\pi)^3}  A_\uparrow \left(\kk, \frac{\kk^2}{2m}+\omega_3-\omega_\text{RF} \right) \nonumber \\
&\quad\quad\quad\quad\quad\quad \times n_F( \frac{\kk^2}{2m}+\omega_3-\omega_\text{RF}). \label{eq:IA2}
\end{align}
Adding the assumptions that we are working at zero temperature and the system has spherical symmetry, we can simplify the expression for the current even further
\begin{align}
I(\omega_\text{RF})=\Omega^2 \int_0^{\sqrt{2m(\omega_\text{RF}-\omega_3)}} \frac{k^2 dk}{2\pi^2}  A_\uparrow(k, \frac{k^2}{2m}+\omega_3-\omega_\text{RF}). \label{eq:IA3}
\end{align}

To apply Eq.~\eqref{eq:IA} to the impurity problem, we separate the spectral function into that of the clean system $A_0(\kk, \omega)$ and corrections that depend on the impurity density $\Delta A(\kk, \omega)$
\begin{align}
A_0(\kk, \omega)=A_0(\kk, \omega)+n_i \left[\Delta A_c(\kk, \omega)+\Delta A_i(\kk, \omega)\right].
\end{align}
Here, we have further separated the impurity contribution $\Delta A(\kk, \omega)=\Delta A_c(\kk, \omega)+\Delta A_i(\kk, \omega)$ into a coherent part that corresponds to the spectral weight of impurity bound states and incoherent part that corresponds to the broadening of the continuum states by impurity scattering. We apply the same criteria to separate the RF transition rate 
\begin{equation} 
I(\omega)=I_0(\omega)+n_i (\Delta I_c(\omega) + \Delta I_i(\omega)), \label{eq:Dci}
\end{equation}
where $I_0(\omega)$ corresponds to the transition rate of a clean system, while $\Delta I_c(\omega)$ and $\Delta I_i(\omega)$ are the coherent and incoherent corrections due to the impurities. 
%\begin{align}
%\mathcal{D}_c(i \omega_n)=\int\frac{d^3\kk}{(2\pi)^3} \frac{1}{\beta}\sum_{i \omega_1}G^0_\uparrow(\kk, i \omega_1)G_3(\kk, i \omega_1 + i \omega_n) \label{eq:Dc}
%\end{align}
%is the convolution of the Green function of the clean system and the $|3\rangle$ state Green function, while
%\begin{align}
%\mathcal{D}_i(i \omega_n)=&\int\frac{d^3\kk}{(2\pi)^3} \frac{1}{\beta}\sum_{i \omega_1}
%\left[G^0(\kk, i \omega_1) T(i \omega) G^0(\kk, i \omega_1)\right]_\uparrow \nonumber \\
%&\quad\quad\quad\quad\quad\quad\quad\quad \times G_3(\kk, i \omega_1 + i \omega_n) \label{eq:Di}
%\end{align}
%is the convolution of the correction to the Green function induced by the impurity and the $|3\rangle$ state Green function.

\begin{figure*}
\includegraphics[scale=0.5]{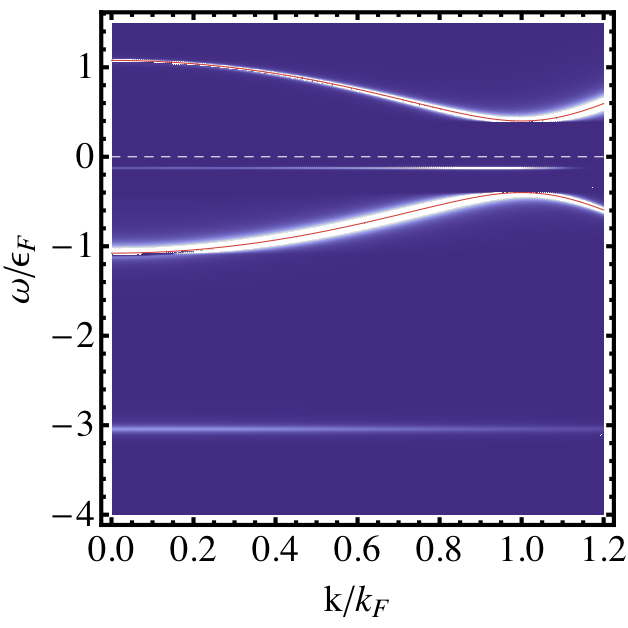} \includegraphics[scale=0.5]{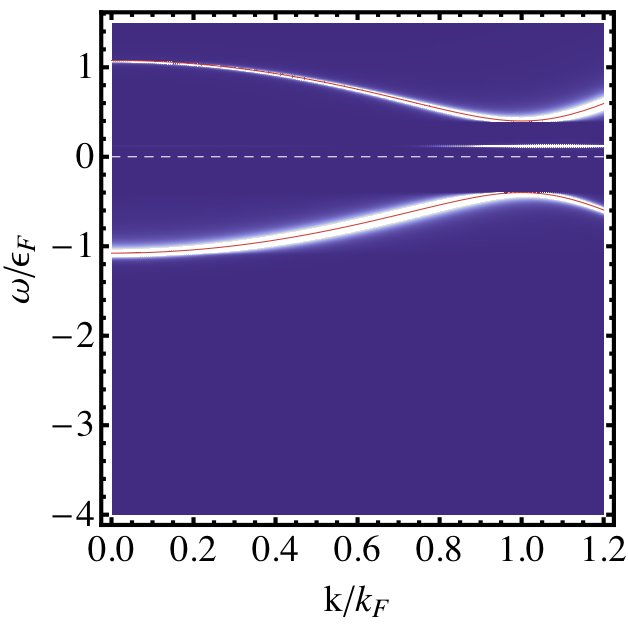}
\caption{Impurity induced correction to the spectral function of the $\upp$ atoms (left) and $\dnn$ atoms (right) as a function of momentum and frequency for the case $a_\uparrow k_F=0.5$, $a_\downarrow=-0.5$, and $\Delta/\epsilon=0.4$ (white -- increase of spectral weight, blue -- no change, red -- decrease). The dashed white line indicates the position of the Fermi energy.  Both $A_{n_i,\uparrow}(\kk,\omega)$ and  $A_{n_i,\downarrow}(\kk, \omega)$  show a depletion of spectral weight along the dispersion curve of the clean system indicated by the red line. $A_{n_i,\uparrow}(\kk,\omega)$ shows an under-sea bound state at $\omega\approx -3 \epsilon_F$ as well as a Shiba state at $\omega \approx -0.13 \epsilon_F$, while $A_{n_i,\downarrow}(\kk, \omega)$ shows only a Shiba state at  $\omega \approx 0.13 \epsilon_F$. In addition, there is spectral weight in the vicinity of the dispersion curve of the clean system which corresponds to impurity induced broadening.   }
\label{fig:A2l}
\end{figure*}

\subsection{RF spectrum of a one-component gas with an impurity}
Suppose that the atom cloud is composed of a single, non-interacting, fermionic species in the hyperfine state $\upp$. To understand the RF induced transition rate, and how it is affected by an impurity, it is useful to begin by describing the spectral function of the $\upp$ fermions. The clean spectral function has the form $A_0(\kk,\omega)=2\pi \delta(\omega-k^2/2m+\epsilon_F)$. The impurity induced corrections to this spectral function $\Delta A(k,\omega)$ are plotted in Fig.~\ref{fig:A1I}a. These corrections move spectral weight away from the clean dispersion and can be separated into an incoherent part that corresponds to the broadening of the continuum band by impurity scattering and a coherent part that corresponds to the impurity bound states. 

In Fig.~\ref{fig:A1I}b, we plot a slice through $\Delta A(k,\omega)$ at fixed $k=0.5k_F$. In the slice we see three main features. First, we see a negative $\delta$-function feature, the location of which coincides with the positive $\delta$-function in $A_0(\kk,\omega)$ (feature 1a). This feature corresponds to the depletion of spectral weight from $A_0(\kk,\omega)$. The spectral weight is transferred in to two regions:  the under-sea bound state, which appears as a positive $\delta$-function in $\Delta A(k,\omega)$ (feature 2);  the spectral weight is also transferred to the vicinity of the negative $\delta$-function feature and corresponds to the broadening of the sharp dispersion of the clean system (feature 1b). Within our classification system, features 1a and 1b correspond to incoherent spectral weight, while feature 2 corresponds to coherent spectral weight. Finally, we point out that although feature 2 is divergent, its frequency integral is finite. Indeed, the full spectral function satisfies the frequency sum rule, which means that the corrections satisfies
\begin{align}
0=\int_{-\infty}^{\infty} \frac{d\omega}{2 \pi} \Delta A(k,\omega),
\end{align}
for all $k$. Having sorted out the spectral function we move on to the question of transition rate.

Since the dispersions of the $\upp$ hyperfine state and $|3\rangle$ state match, the clean part of the transition rate is sharply peaked at $\omega=\omega_3+\epsilon_F$ and has the form
\begin{align}
I_0(\omega_\text{RF})=\Omega^2 \frac{k_F^3}{3 \pi} \delta([\omega_3+\epsilon_F]-\omega_\text{RF}).
\label{eq:1cc}
\end{align}
At this point we pause to make several remarks. First, we remark that we have been following the notation in which the bottom of the $\upp$ band is shifted to the frequency $-\epsilon_F$. Therefore, the frequency difference between the bottom of the $\upp$ band and the bottom of the $| 3 \rangle$ band is $\tilde{\omega}_3=\omega_3+\epsilon_F$.  As a result, the frequency $\tilde{\omega}_3$ and not $\omega_3$ features in the transition rate formula Eq.~\eqref{eq:1cc}.  In the ultracold atom context, it is natural to fix the ``bare" splitting as $\tilde{\omega}_3$ instead of $\omega_3$, since the bottom of the $\upp$ band does not move as the atom density is changed. Our second remark concerns the trapping potential. It is important to focus the RF radiation on the center of the trap in order to avoid the spatial smearing (due to shift of the Fermi energy), as discussed in Ref.~\cite{KetterleZwierleinNotes}. 

%If the trapping potentials for the $\upp$ and $| 3 \rangle$ hyperfine states are different, then there will be a trap induced shift of $\tilde{\omega}_3$ as a function of position in the trap. This shift will result in the smearing of the sharp peak predicted by Eq.~\eqref{eq:1cc} and will hinder the observation of the signal from bound states. 

Next, we come back to the effects of the impurity. For positive scattering length, there is an impurity bound state which results in a coherent correction to the transition rate
\begin{align}
\Delta I_c(\omega_\text{RF})=\Omega^2 \, 2 \frac{\sqrt{2 m a^2 (\omega_\text{RF}-[\omega_3+\epsilon_F])-1}}{m a^2  (\omega_\text{RF}-[\omega_3+\epsilon_F)])^2}. \label{eq:Ic1}
\end{align}
In addition to this coherent correction there is also an incoherent correction, that occurs regardless of the sign of the scattering length, and results in the broadening of the sharp transition rate of the clean state. We plot the impurity induced corrections to the transition rate in Fig.~\ref{fig:graphnormal} for both negative and positive scattering length. For the positive scattering length case, we highlight the coherent part of the transition rate, given by Eq.~\eqref{eq:Ic1}, with pink shading. The incoherent part of the transition rate correction is composed of a negative $\delta$-function feature (indicated by an arrow in Fig.~\ref{fig:graphnormal}) and a broad positive feature. The $\delta$-function feature corresponds to feature 1a discussed above: the depletion of spectral weight (and thus transition rate) from the clean spectral function. On the other hand the broad positive feature corresponds to feature 1b: the broadening of the dispersion curve of the clean system. Since feature 1b is divergent, we cut it off with a wavy line. As discussed above, this divergence is a spurious consequence of the expansion in impurity density, and we do not expect to see it in experiment. 

%Combining the Green function of a single component Fermi gas Eq.~\eqref{eq:1GF} and the T-matrix for the impurity Eq.~\eqref{eq:1T} together with Eqs.~\eqref{eq:I}, \eqref{eq:D}, \eqref{eq:Dci}, \eqref{eq:Dc}, and  \eqref{eq:Di} we obtain the RF induced transition rate 
%\begin{equation} 
%\frac{I(\omega)}{\Omega^2}=\frac{2m k_F}{6\pi}\left[\delta(\omega-\omega_3)\epsilon_F+n_i \frac{8}{\pi k_F a}\frac{\sqrt{\frac{\omega-\omega_3}{\epsilon_F}-\frac{1}{(k_F a)^2}}}{\left(\frac{\omega-\omega_3}{\epsilon_F}\right)^2}\right].
%\end{equation}
%The two terms that appear in the transition rate correspond to exciting "clean" fermions and fermions in the vicinity of the impurity, respectively. We plot the transition rate as a function of RF drive frequency in Fig.~\ref{fig:graphnormal}. We can understand the $\delta$-function in the first term, which is depicted as the blue spike in Fig.~\ref{fig:graphnormal}, as a consequence of the assumptions that the RF drive does not impart momentum on the fermions and the that the dispersions of the $\upp$ and $|3\rangle$ fermions are the same. The impurity breaks translational invariance and thus results in features away from the frequency $\omega=\omega_3+\epsilon_F$. However, the drive frequency must exceed the impurity binding energy $\omega > \omega_3+\omega_b$ resulting in an extended feature with a sharp "absorption edge" depicted by a black line in Fig.~\ref{fig:graphnormal}. 

\subsection{RF spectrum of a BCS gas with an impurity}
In the clean BCS system, the fermion spectral function (for both species of fermions) has the form 
\begin{align}
A_{0,\sigma}(\kk,\omega)=&\frac{\pi}{E_k} \left[(E_k+\xi_k) \delta(\omega-E_k)\right.\nonumber\\
&\hspace{2cm} \left.+(E_k-\xi_k) \delta(\omega+E_k)\right], \label{eq:ABCS}
\end{align} 
where $E_k=\sqrt{\xi_k^2+\Delta^2}$. The main feature of this spectral function is the superconducting gap in the density of states around the Fermi-surface. As before, the action of the impurity is to modify the clean Green functions and consequently the spectral functions.

We begin by investigating how this spectral function is modified by the presence of the impurity atom, i.e. we compute $-2 \Im G_0(\kk,\omega+i0^+) T(\omega+i0^+) G_0(\kk,\omega+i0^+)$. We plot the change in the spectral function for both species of fermions induced by a magnetic impurity having $k_F a_\uparrow =0.5$ and $k_F a_\downarrow =-0.5$ in Fig.~\ref{fig:A2l}. Similar to the case of the single component gas, we see that the impurity has two effects. First, it induces a broadening of the continuum states. Second, it induces the formation of bound states. For the $\upp$ fermions it induces a Shiba state just under the Fermi energy, while for  $\dnn$ fermions it induces a Shiba state just above the Fermi energy. In addition, as $a_\uparrow$ is positive, the impurity induces an under-sea state for the $\upp$ fermions that is analogous to the under-sea state of the one component gas. 

The RF spectrum for the clean BCS system is plotted in Fig.~\ref{fig:graphBCS}a, and the impurity induced corrections for the up and down atoms are plotted in Figs.~\ref{fig:graphBCS}b and \ref{fig:graphBCS}c, respectively. The corrections to the RF spectrum due to the magnetic impurity are strongest for the $\upp$ to $|3\rangle$ transition, depicted in Fig.~\ref{fig:graphBCS}b. These consist of: (1) a dramatic filling of the gap, i.e. transitions to the left of the threshold frequency for the clean system, associated with the Shiba state below the Fermi energy; and (2)  an edge in the spectrum that appears to the right of the main peak for the clean system associated with under-sea bound state. In the next three subsections we give analytical expressions for the RF spectrum of the clean system and the corrections due to under-sea and Shiba bound states.

\subsubsection*{RF spectrum of the clean system} 
Using Eqs.~\eqref{eq:IA} and \eqref{eq:ABCS} we find that the transition rate for the clean system is
\begin{equation} 
I_0(\omega) = \Omega^2 \frac{m^{3/2} \Delta^2 \sqrt{(\omega-\omega_3)^2-\Delta^2-\epsilon_F^2}}{2\pi (\omega-\epsilon_F-\omega_3)^{5/2}}.
\end{equation}
We note that by dividing our expression by the particle density we recover the transition rate per particle established by Ketterle and Zwierlein~\cite{KetterleZwierleinNotes}. 
We plot this transition rate in Fig.~\ref{fig:graphBCS}a. The sharp onset at low frequencies corresponds to exceeding the threshold frequency
\begin{equation} 
\omega_{th} = \omega_3+\sqrt{\epsilon_F^2+\Delta^2}-\epsilon_F,
\end{equation}
associated with the band bottom.

\subsubsection*{Under-sea states} 
We follow the approximations of subsection \ref{sec:undersea}, $\omega\leq-\epsilon_F$, $\Delta \approx 0$, and use the approximate T-matrix of Eq.~\eqref{eq:Tundersea}. Around the pole $\omega_b=-\epsilon_F-\frac{1}{2ma_1^2}$, the T-matrix takes the asymptotic form
\begin{equation} 
T(\omega\approx\omega_d)\approx \frac{1}{\omega-\omega_b}\left(\begin{array}{cc} \frac{2\pi}{a_1 m^2} & 0 \\ 0 & 0 \end{array}\right). \label{}
\end{equation}
We recognize that in the vicinity of the bound state, the singularity of the $[1,1]$ component of the T-matrix has the same form as the singularity of the T-matrix in the single component gas case, Eq.~\eqref{eq:1T}. Thus, within our approximation $\Delta \approx 0$, the coherent part of the RF spectrum due to an under-sea bound state is identical to that of the single component gas, Eq.~\ref{eq:Ic1}.
This contribution is indicated by the pink shaded region on the right of Fig.~\ref{fig:graphBCS}b.

\subsubsection*{Shiba states} 
Following the assumption of subsection~\ref{sec:Shiba} ($|\omega|<\Delta\ll\epsilon_F$) and using the T-matrix of equation~\eqref{Tintragap} we compute the coherent contribution to the RF transition rate from a Shiba bound state. For the coherent contribution we focus solely on the pole located at $\omega=-|\omega_b|$, which exists in either the first or second subspace of the T-matrix depending on which domain of the $\{\frac{1}{a_1},\frac{1}{a_2}\}$ plane we are working, see Fig.~\ref{fig:a1a2plane}. We assume that we are working at sufficiently low temperature so that only the negative frequency Shiba state is filled, and focus on the case of the negative frequency pole being in the first subspace. If it is in the second subspace, then the filled Shiba state corresponds to a $\dnn$ atom, and thus to detect it we must use the RF transition $\dnn \rightarrow |3\rangle$ instead of $\upp \rightarrow |3\rangle$.

\begin{figure}
\includegraphics[width=8cm]{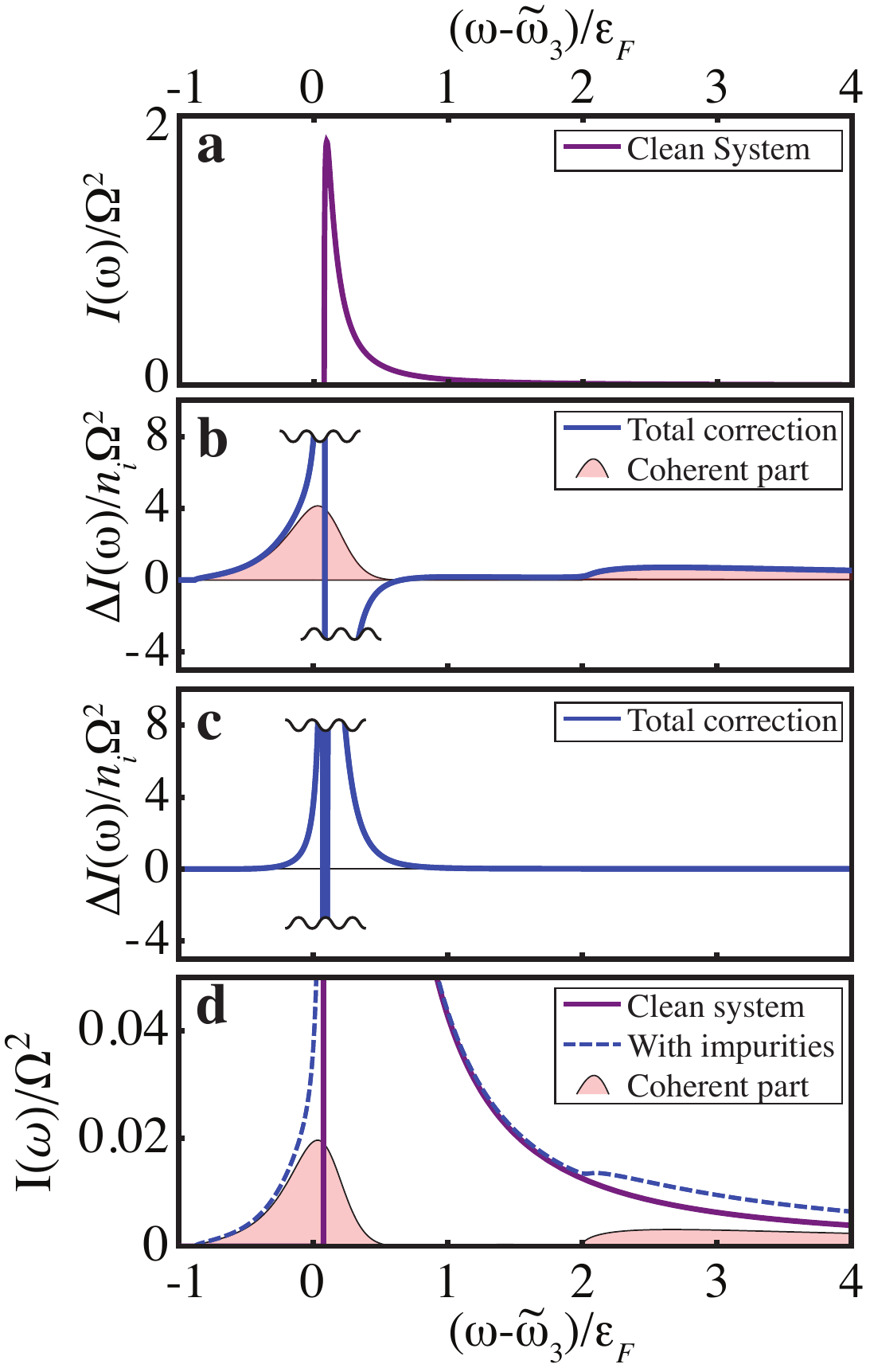}
\caption{(a) RF transition rate for BCS state as a function of the drive frequency $\omega$. Corrections to the transition rate for the $\upp$ atoms (b) and $\dnn$ atoms (c). (d) Total transition rate (clean+corrections) for 10\% concentration of impurities, with divergences smoothed out. Throughout we have used  $a_\uparrow k_F=0.5$, $a_\downarrow=-0.5$, and $\Delta/\epsilon=0.4$.  In (b) the coherent part of the transition rate correction, i.e. the part induced by the Shiba and the under-sea bound states is indicated by pink shading, with the peak on the left corresponding to the Shiba state and the peak on the right to the under-sea state. Similar to the case of the single component Fermi gas, the incoherent part of the transition rate correction is divergent at this order in impurity density (see Fig.~\ref{fig:graphnormal}). Therefore, the total transition rate correction which is plotted in (b) and (c) is also divergent, and we cut it off with wavy lines, as before. }
\label{fig:graphBCS}
\end{figure}

Around the pole $\omega_b$, the asymptotic form of the T-matrix is found to be
\begin{widetext}
\begin{align} 
T(\omega\simeq\omega_b) \simeq &  \frac{2 \pi}{m k_F} \frac{\frac{1}{k_F a_1}-\frac{1}{k_F a_2}}{\left(\frac{1}{k_F a_1}-\frac{1}{k_F a_2}\right)^2+\left(1+\frac{1}{k_F a_1 k_F a_2}\right)^2}\\ 
& \frac{1}{\omega-\omega_b}\left(\begin{array}{cc} \omega_b-\frac{1}{k_F a_2}\sqrt{\Delta^2-\omega_b^2} & -\Delta \\ -\Delta & \omega_b+\frac{1}{k_F a_1}\sqrt{\Delta^2-\omega_b^2} \end{array}\right)=\frac{1}{\omega-\omega_b} R,
\end{align}
\end{widetext}
where we define $R$ to be the regular part of the T-matrix in the vicinity of the pole. The coherent contribution to the spectral function must come from the above pole of the T-matrix. Combining the above form of the T-matrix with the clean BCS Green function Eq.~\ref{eq:GBCS} and the Golden Rule formula Eq.~\ref{eq:Ic1} we obtain
\begin{align}
\Delta I_c(\omega) =\Omega^2 \frac{m k_w}{\pi} \left[ G_0(k_w \omega_b) \cdot R \cdot G_0(k_w, \omega_b) \right]_{11},
\end{align}
where $k_w=\sqrt{ 2 m (\omega+\omega_b-\omega_3)}$ and $\cdot$ indicates a matrix product and $[]_{11}$ indicates the $[1,1]$ component of the matrix. From this expression, we see that for a Shiba state the threshold frequency for RF transition is $\omega_\text{th}=\omega_3-\omega_b$. The coherent contribution of the Shiba state to the RF spectrum is indicated by the pink shaded region on the left of Fig.~\ref{fig:graphBCS}b. From the spectrum we see that most of the weight in the coherent part of the RF spectrum occurs at frequencies significantly higher than $\omega_\text{th}$. This is due to the fact that the Shiba state has most of its spectral weight concentrated at momenta $\sim k_F$.

%\begin{figure}[h] 
%		\includegraphics[scale=0.80]{set_of_spectra2.pdf}
%	\caption{Results for the exact numerical calculation of the RF spectra in presence of an impurity with $k_F=\sqrt{2}$, $\epsilon_F=1$, $\Delta=0.4$, $\omega_3=5$, and for different values of $a_1$ and $a_2$. For each graph, the black curve represents the impurity-free part of the spectrum, the red curve the intra-gap state part, and the dashed blue curve the under-band state part.}
%	\label{fig:set_of_spectra2}
%\end{figure}

\section{Experimental Realization}
\label{sec:experiment}
In this section, we turn to the experimental realization of such a system. In a typical dilute ultracold atomic gas, the Fermi wavevector will be on the order of $k_F \sim 1/4000 a_0$, where $a_0$ is the Bohr radius. The typical order of magnitude of the scattering length, in the absence of Feshbach resonance, is given by the Van der Waals interaction $a \sim 50-100 a_0$. In this regime, the $k_F(a_\uparrow-a_\downarrow)$ and $k_F(a_\uparrow+a_\downarrow)$ amplitudes always remain smaller than unity. Thus, in the absence of the resonance, the Fermion-Impurity (F-I) scattering lengths $a_\uparrow$ and $a_\downarrow$ have roughly the same background values. As a result, the magnetic character of the interaction is vanishingly small and thus the Shiba states are too close to the gap edges to lead to observable results. 

The experimental conditions shall thus be chosen such as these two scattering lengths have widely different values, that is, close to an interspecies Feshbach resonance corresponding to one of the F-I interactions. Simultaneously, we wish to stay close above the F-F Feshbach resonance in order to maintain the large negative value of the associated scattering length. In conclusion, the impurity atom must be chosen to have a Feshbach resonance with one of the fermion hyperfine levels for a magnetic field slightly superior to the F-F resonant value. In addition to the requirement for Feshbach resonances, it is necessary to be able to confine the impurity very tightly in an optical lattice, while the
fermions should still be relatively free. This would favor using a light fermion and a relatively heavy impurity atom, and employing a wavelength for the optical lattice that is near-detuned with respect to the
optical transition of the impurity atom

\begin{figure*}
\begin{minipage}[t]{7cm}
\includegraphics[width=7cm]{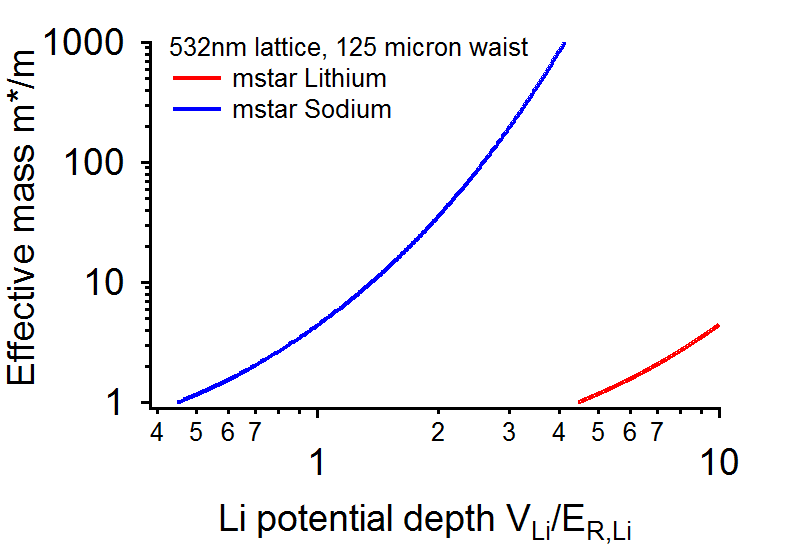} 
\end{minipage}
\begin{minipage}[b]{0.5cm}
{\Large \bf a} 

\vspace{4cm}
\end{minipage}
\hspace{1cm}
\begin{minipage}[t]{7cm}
\includegraphics[width=7cm]{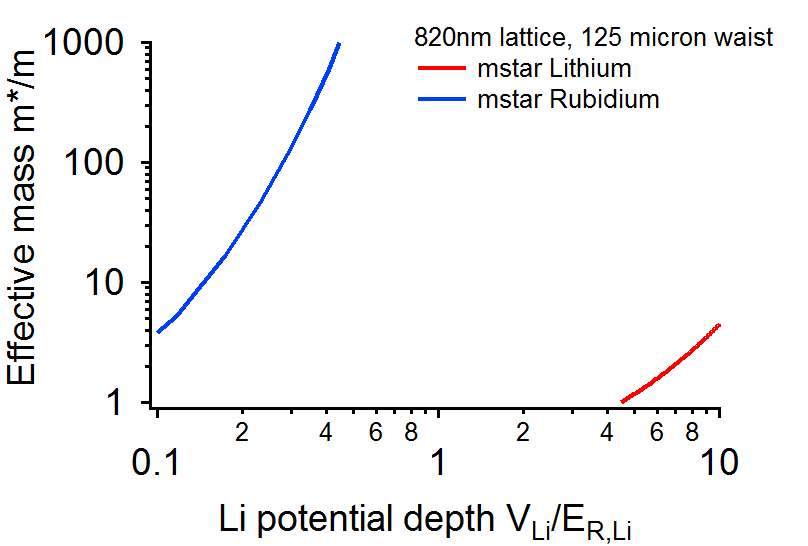}
\end{minipage}
\begin{minipage}[b]{0.5cm}
{\Large \bf b} 

\vspace{4cm}
\end{minipage}
\caption{(a) Effective mass $m^*$ of lithium and sodium atoms in a $532\,\text{nm}$ lattice as a function of the lattice depth (measured in lithium recoil energies). Using a potential depth of $\sim 4\,E_{R,Li}$ it is possible to localize the sodium atoms (that serve as impurities) while lithium atoms remain itinerant.  (b) Effective mass for lithium and rubidium atoms in a $820\,\text{nm}$ lattice. Lattice depths between about $0.5$ and $4\,E_{R,Li}$ can be used to localize the Rb atoms while Li remains itinerant.}
\label{mstar}
\end{figure*}

One possible choice of fermion atoms are the two lowest hyperfine states of $^6$Li, which have a Feshbach resonance at $B_0=834\,\text{G}$. In order to achieve a BCS state, we want a ``slightly superior'' magnetic field, which means here that the difference between $B$ and $B_0$ shall be kept within the range of the Li-Li resonance width, which is approximately $\Delta B\sim 300\,\text{G}$. 
Amongst the few easily trappable bosons or fermions that could form a stable 
ultracold mixture with $^6$Li, the boson $^{23}$Na seems to rather well fit the above condition. 
Several Feshbach resonances have been observed between the $^{23}$Na hyperfine 
ground state and the ground state $|1\rangle$ of $^6$Li, at magnetic fields close to 
the broad $^6$Li-$^6$Li Feshbach resonance [Ref.~\cite{Stan2004}]. 
From these data, Gaesca, Pellegrini and C\^ot\'e deduce in [Ref.~\cite{LiNa}] the existence of further resonances between Na and $^6$Li in states $|1\rangle$ and $|2\rangle$ between $834$ 
and $1500\,\text{G}$. A complete list of predicted resonances is presented by Stan in Refs.~\cite{Stan, Stan2004}, along with a discussion of whether each corresponding hyperfine mixture may or may not be stable towards losses due to spin-exchange collisions. For sodium-lithium mixtures, a possibility is to use a green lattice laser at $532\,\text{nm}$. The effective mass for the sodium and lithium atoms as a function of lattice depth is plotted in Fig.~\ref{mstar}a. For a lattice beam with $\sim120\,\mu\text{m}$ waist, and a potential depth of about 4 lithium recoil energies, the sodium tunneling is essentially switched off (with an effective mass of $m^* = 1000m$), while lithium is still forming an itinerant Fermi sea (with an effective mass of $m^* \sim m$).

Another interesting combination are the lithium-rubidium interspecies resonances that
have been found in Ref.~\cite{Deh}. Here, there is a very interesting resonance at $882\,\text{G}$ which is $1.3\,\text{G}$ wide, not far from the $834\,\text{G}$ resonance in lithium, and --- as required for the assumptions in the paper --- on the BCS side. The advantage of using the $882\,\text{G}$ Rb-Li resonance over any of the Na-Li resonances is technical: the $882\,\text{G}$ Rb-Li resonance has a width of  $1.3\,\text{G}$ while the Na-Li resonances have widths of $\sim 300\,\text{mG}$.
However, depending on the Li density, the $882\,\text{G}$ Rb-Li resonance may lie in the BEC-BCS crossover regime as opposed to the BCS regime. We suspect that the Shiba states will continue into the crossover regime, however determining their properties requires extending our theory. Alternatively, there is a Rb-Li resonance at $1067\,\text{G}$, which is wide ($10.6\,\text{G}$), but lithium is then less strongly interacting,  making it more difficult to attain a superfluid. For lithium-rubidium, one could use a laser tuned to about 820nm (see Fig.~\ref{mstar}b). As rubidium is so heavy compared to lithium, it makes for a very good localized impurity.

\section{Outlook}
\label{sec:outlook}
We suggest that the ``implantation" of magnetic impurities into ultracold atom systems could lead to many exciting possibilities. As already mentioned, one class of possibilities involves leveraging the interaction of magnetism and superconductivity. This class includes the application of magnetic impurities as local probes,  which is the subject of the present paper. Another possibility is to study how the pair-breaking effect of the magnetic impurities leads to the destruction of superconductivity under various conditions. In 3D one would hope to realize the transition from gap-full to gapless superconductivity. On the other hand, in 2D and 1D the pair-breaking effect of the magnetic impurities is  predicted to drive the superconductor-insulator transition.

Another class of possibilities involves the Kondo effect. We already see a precursor to the Kondo level in the under-sea bound state. The nature of this under-sea bound state should undergo a dramatic transformation as we turn on the Kondo effect by changing the fermion-fermion interactions from attractive to repulsive. Significantly, using an optical lattice to localize the impurity atoms naturally invites the experimental realization of the Kondo-lattice model in the setting of ultracold atoms. The Kondo-lattice model is, in turn, a stepping stone on the path of studying itinerant magnetism. 

One significant difficulty in seeing magnetism in the setting of ultracold atoms has been the issue of achieving sufficiently low temperature. Perhaps magnetism without an underlying lattice could be technically advantageous. That is perhaps it will be easier to achieving the Kondo temperature by avoiding the lattice induced losses that feature prominently in the quest to achieve a magnetic transition (e.g. N\'eel temperature) in lattice systems.

\section{Conclusions}
\label{sec:conclusions}
We have investigated the possibility of introducing a magnetic impurity into a cloud of ultracold fermions. In particular we have focused on the realization of a localized impurity atom that is immersed in a one or two component Fermi-gas. To understand the action of the impurity atom on the fermions, we have argued that it can be described by an effective scattering length, at least for the case of a broad resonance with a sufficiently tight impurity confining potential, which we relate to the impurity-fermion scattering length in vacuum. 

Using the effective scattering length description, we find the effects of the impurity on the free Fermi gas as well as a two component BCS condensate. In both cases we find that if there is a positive effective scattering length then the impurity forms an ``under-sea" bound state. In addition impurity scattering breaks translational invariance and thereby broadens the spectral function of the clean system. Finally, for the BCS state if the impurity-fermion scattering lengths are different then the impurity always induces a pair of Shiba bound states inside the gap of the superconductor.

We demonstrate that the impurity bound states appear as additional features in Radio Frequency spectroscopy that should be detectable experimentally. Specifically, we suggest that $^6$Li BCS condensate with $^{23}$Na impurities could be a potentially fruitful experimental system for study of magnetic impurities. We speculate that beyond the study of bound states of dilute impurities, the same setup in combination with RF spectroscopy could be useful for study of gapless superconductivity,  Kondo effect, Kondo lattices, and other problems that combine localized moments and itinerant fermions.

\acknowledgements
The authors acknowledge support from a grant from the Army Research Office with funding from the DARPA OLE program, CUA, NSF Grant No. DMR-07-05472 and PHY-06-53514, AFOSR-MURI, the AFOSR Young Investigator Program, the ARO-MURI on Atomtronics, and the Alfred P. Sloan Foundation.
%\end{acknowledgements}

\appendix
\section{Scattering Problem}
\label{app:Scattering}
In this appendix, we state the scattering problem for the case of a confined impurity, and provide an analytic solution under special circumstances. We describe the scattering of a single fermion off of the trapped impurity by the Hamiltonian
\begin{equation}
H=\frac{p_i^2}{2 m_i} + \frac{p_\alpha^2}{2 m_\alpha} + V_{i-\alpha}(r_i-r_\alpha) + \frac{1}{2} m_i \omega_t^2 r_i^2,
\label{eq:SH}
\end{equation}
where $p_i$, $r_i$ and $m_i$ stand for the momentum, position, and mass of the impurity, $p_\alpha$, $r_\alpha$, and $m_\alpha$ for the momentum, position, and mass of the scattering fermion, $V_{i-\alpha}$ is a pseudo potential that describes scattering of the fermion off of the impurity in vacuum. Eq.~\eqref{eq:SH} completely defines the scattering problem. However, in general, the equation must be solved numerically. 

\begin{figure}
\includegraphics[width=7cm]{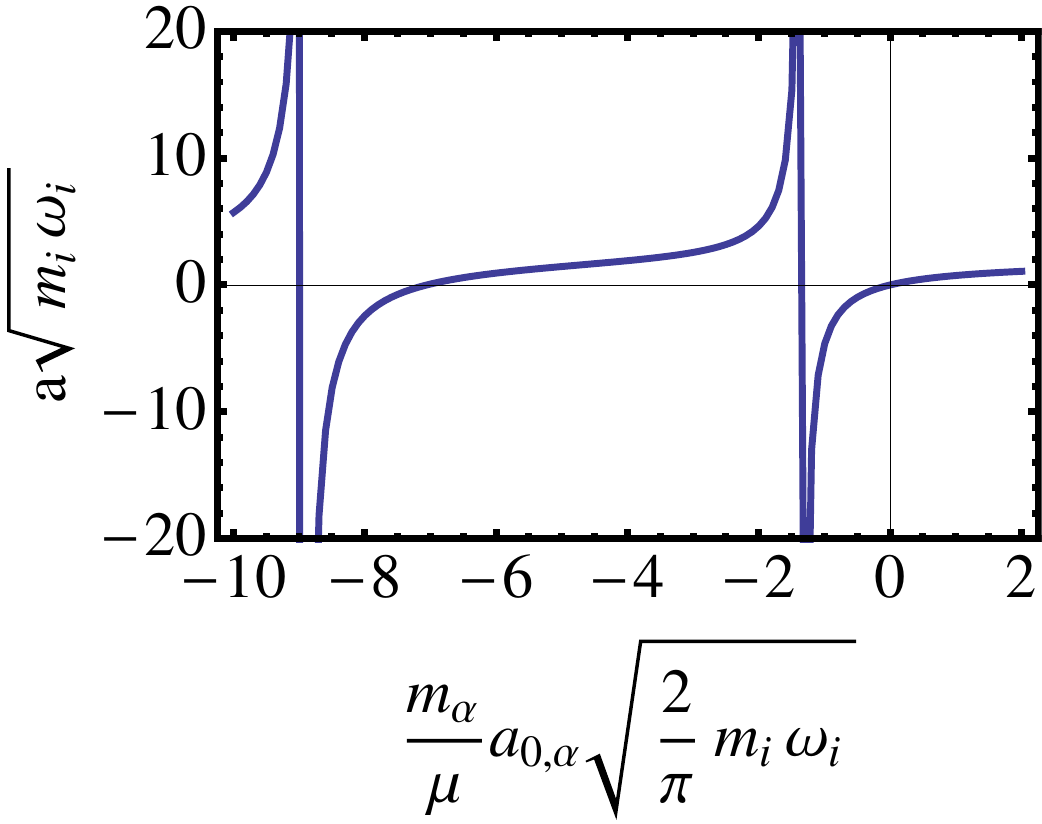}
\caption{Effective scattering length $a$ that describes the scattering of a free fermion of mass $m_\alpha$ on an impurity of mass $m_i$ localized in a harmonic potential of frequency $\omega_i$ as a function of fermion -- impurity atom interaction strength (scattering length in vacuum $a_{0,\alpha}$) computed in the Frozen impurity approximation. For small $a_{0,\alpha}$, $a$ depends linearly on $a_{0,\alpha}$, Eq.~\eqref{eq:LSA}. However,  for large $a$ we see a deviation from linear law. For large negative $a$ it is possible to form bound states of the fermion, which result in Feshbach resonances at $(m_\alpha a_{0,\alpha} / \mu) \sqrt{2 m_i\omega_i/\pi} \approx \{-1.5, -9, ... \}$.}
\label{fig:resonance}
\end{figure}

For the sake of achieving a quantitate answer, we make additional assumptions. First, we assume that the effective range of the pseudo potential $V_{i-\alpha}$ is much narrower than the spatial extent of the harmonic oscillator ground state $\sqrt{\hbar/m \omega_i}$. Combining this assumption with the assumption that the typical collision energy scales $\hbar \omega_i$ are much smaller than the characteristic resonance scale $\frac{\hbar^2}{\mu a_{0,\alpha}^2}$, we can replace the interaction potential by a $\delta$-function $V_{i-\alpha}(r_i-r_\alpha)=\frac{2 \pi a_{0,\alpha}}{\mu} \delta(r_i-r_\alpha)$.

Finally, to obtain an analytical answer we make the frozen impurity orbital approximation. That is we first assume that we can write the wavefunction for the fermion and the impurity in a product form
\begin{align}
\Psi(r_i,r_\alpha)=\psi_i(r_i) \psi_\alpha(r_\alpha),
\end{align}
and then we assume that $\psi_i(r_i)$ is frozen to be the impurity ground state wavefunction.  This approximation is similar in spirit to the Born-Oppenheimer approximation that a heavy fermion is moving in the field of a fast impurity, with the additional assumption that the interaction between the two is sufficiently small that the impurity wavefunction is only weakly effected by the fermion. The approximation is valid for $a_{0,\alpha} \ll (\mu/m_i) \sqrt{\hbar/m_i\omega_i}$. The product wavefunction and the frozen impurity wavefunction approximations often appear in scattering theory. A particularly analogous problem where these approximations have been extensively used is the elastic scattering of a low energy electron from a hydrogen atom~\cite{Schwartz}. In this example, the role of the confinement potential is played by the electrostatic potential of the nucleus, which serves to localize the electron of the hydrogen atom. Within the frozen impurity wavefunction approximation, we define the effective potential that the fermion feels to be
\begin{align}
V_\text{eff}(r)= \frac{2 \pi a_{0,\alpha}}{\mu} |\psi_0(r)|^2
\end{align}
where $\psi_0(r)=(m_i \omega_i / \pi \hbar)^{3/4} e^{-m \omega_i r^2/2\hbar}$ is the ground state wavefunction of the impurity. The effective scattering length, for small $a_{0,\alpha}$, is given by
\begin{align}
a_\alpha=\frac{2.12 \, a_{0,\alpha} m_\alpha}{\mu}\left[1+O\left(\frac{2\, a_{0,\alpha} m_\alpha \sqrt{m_i \omega_i}}{\mu}\right)\right].
\label{eq:LSA}
\end{align}

We note that even within our simple approximation, we find ``geometric" resonances that are induced by bound states of $V_\text{eff}$. These resonances can be clearly seen in the plot of the effective scattering length as a function of $a_{0,\alpha}$ in Fig.~\ref{fig:resonance}. We expect that some of these resonance would survive in a more complete theory of the scattering process, and could be used in experiment to tune the ``magnetism" of the impurity.


\begin{thebibliography}{99}
\bibitem{Shiba} H. Shiba, Prog. Theor. Phys. \textbf{40}, 435 (1968).

\bibitem{AGD} A. A. Abrikosov, L. P. Gorkov, I. E. Dzyaloshinski, {\it Methods of quantum field theory in statistical physics} (Dover Publications, New York, 1975).

\bibitem{Popov} V. N. Popov, Zh. Eksp. Teor. Fiz. {\bf 50}, 1550 (1966) [Sov. Phys. JETP {\bf 23}, 1034 (1968)].
\bibitem{Keldysh} L. V. Keldysh and A. N. Kozlov, Zh. Eksp. Teor. Fiz. {\bf 54}, 978 (1968) [Sov. Phys. JETP {\bf 27}, 521 (1968)].
\bibitem{Eagles} D. M. Eagles, Phys. Rev. {\bf 186}, 456 (1969).
\bibitem{Leggett} A. J. Leggett, J. Phys. (Paris), Colloq. 41, 7 (1980).
\bibitem{Jin} C. A. Regal, M. Greiner, and D. S. Jin, Phys. Rev. Lett. {\bf 92}, 040403 (2004).
\bibitem{Bartenstein} M. Bartenstein, A. Altmeyer, S. Riedl, S. Jochim, C. Chin, J. Hecker-Denschlag, and R. Grimm, Phys. Rev. Lett. {\bf 92},  120401 (2004).
\bibitem{Zwierlein2004} M. W. Zwierlein, C. A. Stan, C. H. Schunck,  S. M. F. Raupach, A. J. Kerman,
 and W. Ketterle, Phys. Rev. Lett. {\bf 92}, 120403 (2004).
\bibitem{Kinast} J. Kinast, S. L. Hemmer, M. E. Gehm, A. Turlapov, and J. E. Thomas, Phys. Rev. Lett. {\bf 92} 150402 (2004).
\bibitem{Bourdel} T. Bourdel, L. Khaykovich, J. Cubizolles, J. Zhang, F. Chevy, M. Teichmann,
L. Tarruell,  S. J. J. M. F. Kokkelmans, and C. Salomon, Phys. Rev. Lett. {\bf 93}, 050401 (2004).
\bibitem{Zwierlein2005} M. W. Zwierlein, J. R. Abo-Shaeer,  A. Schirotzek, C. H. Schunck, and
W. Ketterle, Nature {\bf 435} 1047 (2005).
\bibitem{Burovski} E. Burovski, E. Kozik, N. Prokof'ev, B. Svistunov, M. Troyer, Phys. Rev. Lett. {\bf 101}, 090402 (2008).
\bibitem{Stringari} S. Giorgini, L. P. Pitaevskii, and S. Stringari, Rev. Mod. Phys. {\bf 80}, 1215Ð1274 (2008).

\bibitem{Ketterle} M. W. Zwierlein, A. Schirotzek, C. H. Schunck, W. Ketterle, 
Science {\bf 311}, 492 (2006).
\bibitem{Hulet} G. B. Partridge, W. Li, R. I. Kamar, Y.-a. Liao, and R. G. Hulet,
Science {\bf 311}, 503 (2006).

\bibitem{Barankov} R. A. Barankov and L. S. Levitov, Phys. Rev. Lett. {\bf 96}, 230403 (2006).
\bibitem{Kehrein} M. Moeckel and S. Kehrein, Phys. Rev. Lett. {\bf 100}, 175702 (2008).

\bibitem{Davis} S. H. Pan, E. W. Hudson, K. M. Lang, H. Eisaki, S. Uchida, and J. C. Davis,
Nature {\bf 403}, 746 (2000).
\bibitem{BalatskyViews} A. V. Balatsky, Nature \textbf{403}, 717 (2000).
\bibitem{Hoffman} J. E. Hoffman, K. McElroy, D.-H. Lee, K. M. Lang, H. Eisaki, S. Uchida, and J. C. Davis,
Science {\bf 297}, 1148 (2002).

\bibitem{Spectroscopy} Y. Shin, C. H. Schunck, A. Schirotzek, and W. Ketterle, Phys. Rev. Lett. {\bf 99}, 090403 (2007).
\bibitem{FermiPolaron} A. Schirotzek, C.-H. Wu, A. Sommer, and M. W. Zwierlein, Phys. Rev. Lett. {\bf 102}, 230402 (2009).

\bibitem{Baym} G. Baym, C. J. Pethick, Z. Yu, and M. W. Zwierlein, Phys. Rev. Lett. {\bf 99}, 190407 (2007). 

\bibitem{Schirotzek} A. Schirotzek, Y. Shin, C. H. Schunck, and W. Ketterle, Phys. Rev. Lett. {\bf 101}, 140403 (2008).

\bibitem{Stewart} J. T. Stewart, J. P. Gaebler, and D. S. Jin, Nature {\bf 454}, 744 (2008).

\bibitem{Castin} P. Massignan and Y. Castin, Phys. Rev. A {\bf 74}, 013616 (2006).

\bibitem{Mahan} G. D. Mahan, {\it Many-Particle Physics}, 3rd ed. (Kluwer Academic/Plenum Publishers, New York, 1981).

\bibitem{KetterleZwierleinNotes} W. Ketterle and M. Zwierlein, {\it Making, probing and understanding ultracold Fermi gases}, in {\it Ultracold Fermi Gases, Proceedings of the International School of Physics Enrico Fermi},
Course CLXIV, Varenna, 20 - 30 June 2006, eds. M. Inguscio, W. Ketterle, and C. Salomon
(IOS Press, Amsterdam, 2008).

\bibitem{Marini} M. Marini, F. Pistolesi, and G.C. Strinati, Eur. Phys. J. B, {\bf 1}, 151 (1998).

\bibitem{Nambu} Y. Nambu, Phys. Rev. {\bf 117}, 648Ð663 (1960).
\bibitem{Ambegaokar} V. Ambegaokar and A. Griffin, Phys. Rev., {\bf 137}, A1151 (1965).

%\bibitem{Aalkola} M.I. Salkola, A.V. Balatsky and J.R. Schrieffer, Phys. Rev. B, {\bf 55}, 1248 (1997).

\bibitem{Stan2004} C. A. Stan, M. W. Zwierlein, C. H. Schunck, S. M. F. Raupach, and W.
Ketterle, Phys. Rev. Lett. {\bf 93}, 143001 (2004).

\bibitem{LiNa} M. Gaesca, P. Pellegrini, R. Côté, Phys. Rev. A, {\bf 78}, 010701(R) (2008).

\bibitem{Stan} PhD Thesis manuscript of C.A. Stan, {\it Experiments with Interacting Bose and Fermi
Gases} (2005).

\bibitem{Deh} B. Deh, C. Marzok, C. Zimmermann, and Ph. W.
Courteille, Phys. Rev. A {\bf 77}, 010701(R) (2008).

\bibitem{Schwartz} C. Schwartz, Phys. Rev. {\bf 124}, 1468 (1961).

\end{thebibliography}
\end{document}